\documentclass[useAMS,usenatbib]{mn2e}

\usepackage{apjfonts}
\usepackage{graphicx}
\usepackage{amsmath}
\usepackage{amssymb}
\title[Relativistic Blast Wave Encounters a Density Jump]{Smooth Light Curves from a
Bumpy Ride: Relativistic Blast Wave Encounters a Density Jump}

\author[E. Nakar and J. Granot]{Ehud Nakar$^{1}$ and Jonathan
Granot$^{2}$\\
$^{1}$Theoretical Astrophysics, Caltech, Pasadena, CA 91125, USA ; udini@tapir.caltech.edu\\
$^{2}$KIPAC, Stanford University, P.O. Box 20450, MS 29, Stanford,
CA 94309, UAS; granot@slac.stanford.edu}

\begin{document}

\maketitle

\begin{abstract}

Some gamma-ray burst (GRB) afterglow light curves show significant
variability, which often includes episodes of rebrightening. Such
temporal variability had been attributed in several cases to large
fluctuations in the external density, or density "bumps". Here we
carefully examine the effect of a sharp increase in the external
density on the afterglow light curve by considering, for the first
time, a full treatment of both the hydrodynamic evolution and the
radiation in this scenario. To this end we develop a semi-analytic
model for the light curve and carry out several elaborate numerical
simulations using a one dimensional hydrodynamic code together with
a synchrotron radiation code. Two spherically symmetric cases are
explored in detail -- a density jump in a uniform external medium,
and a wind termination shock. The effect of density clumps is also
constrained. Contrary to previous works, we find that even a very
sharp (modeled as a step function) and large (by a factor of $a \gg
1$) increase in the external density does not produce sharp features
in the light curve, and cannot account for significant temporal
variability in GRB afterglows. For a wind termination shock, the
light curve smoothly transitions between the asymptotic power laws
over about one decade in time, and there is no rebrightening in the
optical or X-rays that could serve as a clear observational
signature. For a sharp jump in a uniform density profile we find
that the maximal deviation $\Delta \alpha_{\rm max}$ of the temporal
decay index $\alpha$ from its asymptotic value (at early and late
times), is bounded (e.g, $\Delta \alpha_{\rm max} < 0.4$ for $a =
10$); $\Delta \alpha_{\rm max}$ slowly increases with $a$,
converging to $\Delta \alpha_{\rm max} \approx 1$ at very large $a$
values. Therefore, no optical rebrightening is expected in this case
as well. In the X-rays, while the asymptotic flux is unaffected by
the density jump, the fluctuations in $\alpha$ are found to be
comparable to those in the optical. Finally, we discuss the
implications of our results for the origin of the observed
fluctuations in several GRB afterglows.

\end{abstract}


\section{Introduction}
\label{sec:intro}

Gamma-ray bursts (GRBs) are produced by a relativistic outflow from
a compact source. The outflow sweeps up the external medium and
drives a strong relativistic shock into it. Eventually the outflow
is decelerated (by $pdV$ work across the contact discontinuity that
separates the ejecta and the shocked external medium) and most of
the kinetic energy is transferred to the shocked external medium
(for recent reviews see \citealt{Piran05,Meszaros06}). The shocked
external medium produces long lived afterglow emission that is
detected in the X-rays, optical, and radio for days, weeks, and
months, respectively, after the GRB.  The afterglow emission is
thought to be predominantly synchrotron radiation. This is supported
both by the broad band spectrum and by the detection of linear
polarization at the level of a few percent in the optical (or near
infrared) afterglow of several GRBs \citep[see][and references
therein]{Covino03}. Inverse-Compton scattering of the synchrotron
photons might dominate the observed flux in the X-rays in some cases
\citep{PK00,SE01,Harrison01}.

In the pre-{\it Swift} era the best monitoring of GRB afterglow
light curves was, by far, in the optical. Most afterglow light
curves showed a smooth power law decay
\citep{Stanek99,LS03,Gorosabel06}, and often also a smooth
achromatic transition to a steeper power law decay that is
attributed to the outflow being collimated into a narrow jet
\citep{Rhoads99,SPH99}. It has been argued \citep{WL00} that the
smoothness of the afterglow light curve is directly related to (and
thus enables to constrain) the smoothness of the external density
field. Nevertheless, some optical afterglows have shown significant
temporal variability, with strong deviations from the more typical
smooth power law behavior. The best examples are GRBs 021004
\citep{Pandey02,Fox03,Bersier03} and 030329
\citep{Matheson03,Sato03,Uemura04,Lipkin04}.

Possible explanations for such temporal variability in GRB afterglow
light curves include variations in the external density
\citep{WL00,Lazzati02,NPG03}, or in the energy of the afterglow
shock. The latter includes energy injection by ``refreshed shocks'' --
slower shells of ejecta that catch up with the afterglow shock on long
time scales \citep{RM98,KP00a,SM00,R-RMR01,GNP03} or a ``patchy
shell'' -- angular inhomogeneities within the outflow
\citep{KP00b,NPG03,HP03,NO04}. Another possible cause for variability
in the afterglow light curve, although it is expected to be quite
rare, is microlensing by an intervening star in a galaxy that happens
to be close to our line of sight. GRB~000301C exhibited an achromatic
bump in its optical to NIR light curves that peaked after $\sim
4\;$days \citep{Sagar00,Berger00} which had been interpreted as such a
microlensing event \citep{GLS00,GGL01,BH05}, although other
interpretations such as a bump in the external density have also been
suggested \citep{Berger00}.

Recent observations by {\it Swift} have found flares in the early
X-ray afterglows of many GRBs
\citep{Burrows05,Nousek06,Falcone06,Obrien06} which are probably due
to late time activity of the central source \citep{Nousek06,Zhang06}.
Early optical variability also appears to be more common than
previously thought \citep[e.g.,][]{Stanek06}, although it is not yet
clear if it is caused by similar mechanisms as the late time optical
variability that had been detected before {\it Swift}.

The most natural forms of variations in the external density are
either clumps on top of a smooth background density distribution, or
a global abrupt change in density with radius. The latter can be,
e.g., the termination shock of the wind from the massive star
progenitor of a long-soft GRB \citep{Wijers01}. Such a stellar wind
environment may have a richer structure and can include an abrupt
increase in density with radius at the contact discontinuity between
shocked wind from two different evolutionary stages of the
progenitor star, as well as clumps that are formed due to
Rayleigh-Taylor instability \citep{R-R05,EGDM06}\footnote{
\citet{R-R05} find that the clump formation may also involve the
Vishniac instability, and once such clumps are formed they stand a
reasonable chance to survive until the time of the core collapse of
the progenitor star.}. Density clumps with a mild density contrast
may also be formed due to turbulence in the external medium.
Furthermore, the external density profile is expected to vary
between different progenitor models \citep{FRY06}.  The variability
that had been observed in optical afterglows was attributed to
density clumps in the external medium \citep{Lazzati02,NPG03}. The
expected observational signatures of the afterglow shock running
into the wind termination shock of the massive star progenitor have
also been considered \citep{Wijers01,R-R01,R-R05,EGDM06,PW06}. In
all these cases it has been argued that there would be a clear
observational signature in the form of a rebrightening in the
afterglow light curve, before approaching the new shallower decay
slope corresponding to the uniform density of the shocked wind.

Here we revisit the effect of density fluctuations on the afterglow
light curve by solving in detail the case of a spherically symmetric
external density with a single density jump (by a factor of $a > 1$)
at some radius $R_0$, while the density at smaller and larger radii
is a (generally different) power law in radius. This is done by
constructing a semi-analytic model for the observed flux due to
synchrotron emission at different power law segments of the spectrum
and by carrying out numerical simulations. The semi-analytic model
takes into account the effect of the reverse shock on the
hydrodynamics and on the emissivity, as well as the effect of the
spherical geometry on the arrival time of photons to the observer.
Being semi-analytic, however, this model uses some approximations
for the hydrodynamic evolution and the resulting radiation.
Therefore, we also perform numerical simulations in which the light
curves are calculated using a hydrodynamic+radiation numerical code.
This code self-consistently calculates the radiation and evolves the
electron distribution in every fluid element, which corresponds to a
computational cell of the one dimensional Lagrangian hydrodynamic
code. The results of this code are used to obtain light curves in
cases of special interest and near spectral break frequencies, as
well as to verify the quality of the semi-analytic model, which is
found to agree well with the numerical results. These calculations
are much more accurate than those presented in previous works, and
our results are significantly different. In all cases we find a very
smooth transition to the new asymptotic power law, with no
rebrightening for an initially decaying light curve.

In \S \ref{sec:spherical} we develop the semi-analytic model. First,
in \S \ref{sec:hydro} a simple analytic model is constructed for the
hydrodynamics, which agrees very well with our numerical results.
Then, in \S \ref{sec:rad} we construct a semi-analytic model for the
observed flux density. Specific case studies (\S \ref{sec: case
study}) are then analyzed in detail, for a wind termination shock (\S
\ref{sec:wind}) and for a spherical density jump in a uniform medium
(\S \ref{sec:uniform}). The light curves for these cases are also
calculated using the numerical code (described in Appendix
\ref{sec:code}). The effect of proximity to a break frequency around
the time of the density jump is investigated in \S \ref{sec:nu_break},
and our main conclusions are found to remain valid also in the
vicinity of the break frequencies. \S \ref{sec:clump} is devoted to a
discussion of the expected observational signatures of density clumps
on top of a smooth underlying external density distribution. Such
density clumps are found to have a very weak observational signature
which would be very hard to detect. In \S \ref{sec:GRBs} we discuss
the implications of our results for the origin of the observed
fluctuations in several GRB afterglows. Our conclusions are discussed
in \S \ref{sec:dis}.

\section{Semi-Analytic model for a Spherical Jump in the External Density}
\label{sec:spherical}

In this section we model a spherical relativistic blast wave that
propagates into a power-law external density profile ($\rho_{\rm
ext} = Ar^{-k}$ with $k<3$) which has a single sharp density jump
(by a factor $a > 1$) at some radius $r = R_0$. The power law index,
$k$, of the external density is allowed to be different at $r < R_0$
($k_0$) and at $r > R_0$ ($k_1$).

The hydrodynamic evolution, as well as the resulting contribution to
the light curves, can be roughly separated into three phases
corresponding to the following ranges of the radius $R$ of the forward
shock: (i) at $R < R_0$ the blast-wave follows a self-similar
evolution \citep[][BM hereafter]{BM76}, (ii) at $R = R_0$ a reverse
shock forms which crosses most of the shell of previously shocked
material (that had been swept up at $r < R_0$) at $R = R_1$, while the
forward shock continues ahead of the density jump but with a reduced
Lorentz factor, (iii) at $R > R_1$ the forward shock relaxes into a
new self-similar evolution corresponding to the new density profile at
$r > R_0$. In the following we first approximate the hydrodynamic
evolution of the different shocks during these three phases and then
calculate the resulting light curves.

\subsection{Hydrodynamics}
\label{sec:hydro}

Consider a spherical ultra-relativistic blast wave (identified with
the afterglow shock), which is well described by the self-similar BM
solution at $R < R_0$, that propagates into the following external
density profile: \\
\begin{equation}\label{EQ rho_ext}
\rho_{\rm ext} = \left\{\begin{matrix}A_0r^{-k_0} & \ \ r<R_0\
,\cr\cr A_1r^{-k_1} & \ \ r>R_0\ .\end{matrix} \right.
\end{equation}
\\
The amplitude of the density jump, i.e. the factor by which the
density increases at $r=R_0$, is given by
\\
\begin{equation}
a\equiv\lim_{\epsilon\to 0}\frac{\rho_{\rm ext}[(1+\epsilon)R_0]}
{\rho_{\rm ext}[(1-\epsilon)R_0]} = \frac{A_1}{A_0}R_0^{k_0-k_1}\ ,
\end{equation}
\\ and is assumed to be larger than unity. The afterglow shock
encounters the jump in the external density profile at a lab frame
time $t_0 = [1+1/2(4-k_0)\Gamma_4^2]R_0/c \approx R_0/c$, where $c$ is
the speed of light ,$\Gamma_4$ is the Lorentz factor of the shock
front just before it encounters the density bump at $r = R_0$, and the
corresponding Lorentz factor of the fluid just behind the shock is
denoted by $\gamma_4 = \Gamma_4/\sqrt{2}$. At $t < t_0$ there are
three regions: the region behind the afterglow shock (subscript `4')
is described by the BM solution with $(A,k) = (A_0,k_0)$, and the two
regions of cold unperturbed external medium (subscripts `0' and `1' at
$r < R_0$ and $r > R_0$, respectively).

When the afterglow shock encounters the jump in the external density
a reverse shock is driven into the hot BM shell, while a forward
shock propagates into the cold higher density external medium at $r
> R_0$. At this stage region `0' no longer exists, but two new
regions are formed so that altogether there are four regions: (i)
the cold unperturbed external medium ahead of the forward shock with
a density $\rho_{\rm ext} = A_1r^{-k_1}$, (ii) the shocked external
medium originating from $r > R_0$, (iii) the portion of the BM shell
that has been shocked by the reverse shock (corresponding to doubly
shocked external medium originating at $r < R_0$), and (iv) the
unperturbed portion of the BM solution which has not yet been
shocked by the reverse shock (i.e. singly shocked external medium
originating from $r < R_0$). These regions are denoted by subscripts
`1' through `4', respectively. Regions 2 and 3 are separated by a
contact discontinuity.

Immediately after $t_0$ [or more precisely, at $0<(t-t_0)/t_0\ll
a^{-1/2}$] the reverse shock reaches only a very small part of the
BM profile (just behind the contact discontinuity) which corresponds
to values $\chi-1\ll 1$ of the self similar variable, $\chi$
\citep[defined in][]{BM76}, so that at this early stage the
conditions in this region may be approximated as being constant with
the values just behind the shock for the BM profile with $(A,k) =
(A_0,k_0)$ at $t = t_0$ (i.e. when the shock radius is $R = R_0$).
Since we are interested in a small range in radius, $\Delta R \ll
R_0$, we can use a planar geometry and solve the relevant Riemann
problem. Region 4 is described by the BM solution while in region 1
we have $\rho_1 = w_1/c^2 = n_1m_p = A_1r^{-k_1}$, $p_1 = e_1 = 0$
and $\gamma_1 = 1$, where $m_p$ is the proton mass.  The pressure
$p$, internal energy density $e$, enthalpy density $w$, rest mass
density $\rho$, and number density $n$ are measured in the proper
frame (i.e. the fluid rest frame). We consider a relativistic
afterglow shock, $\gamma_4 \gg 1$, and a density contrast which is
not too large such that even after the afterglow shock crosses the
density bump it will still be relativistic (i.e. $\gamma_4 \gg \psi
\sim a^{1/4}$, see equation [\ref{EQ_psi}]). Under these conditions,
in regions 2 and 3 the fluid is relativistically hot, $\rho_2c^2 \ll
p_2$ and $\rho_3c^2 \ll p_3$. Therefore, the adiabatic index in
regions 2, 3 and 4 is $4/3$, implying $p_i = e_i/3 = w_i/4$ in these
regions.  This leaves eight unknown quantities: $\gamma$, $n$ and
$e$ in regions 2 and 3, as well as the Lorentz factors of the
reverse shock, $\Gamma_r$, and of the forward shock, $\Gamma_f$.
Correspondingly, there are eight constraints: three from the shock
jump conditions at each of the two shocks, and two at the contact
discontinuity: $e_2 = e_3$ and $\gamma_2 = \gamma_3$. The shock jump
conditions simply state the conservation of energy, momentum, and
particle number across the shock, which is equivalent to the
continuity of their corresponding fluxes. At the rest frame of the
shock front they correspond to the continuity of $w\gamma^2v$,
$w\gamma^2(v/c)^2+p$, and $n\gamma v$, respectively, across the
shock (where $v$ is the fluid velocity measured in that frame, while
$p$, $n$, and $w$ are measured in the rest frame of the fluid).
Unless stated otherwise, all velocities and Lorentz factors are
measured in the rest frame of the unperturbed external medium, which
is identified with the lab frame where the flow is spherical.

\begin{figure}
\vspace{-0.28cm}
\hspace{-0.05cm}
\includegraphics[width=8.5cm]{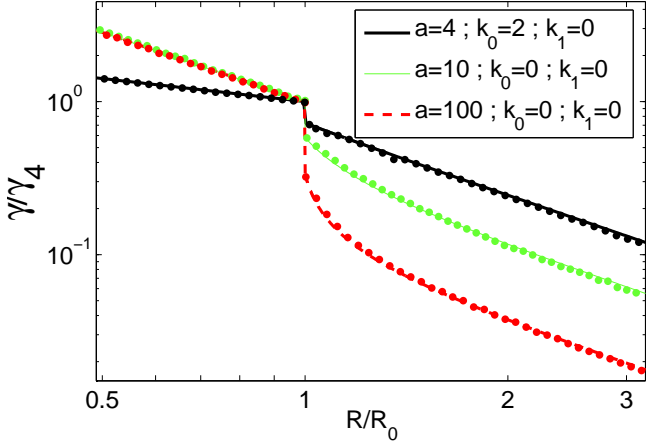}
\caption{\label{FIG R_Gamma} The Lorentz factor of the forward shock
as function of radius for three different density profiles that are
described by Eq. (\ref{EQ rho_ext})\ \ (see legend for the parameters
of each profile). The solid lines show our analytic approximation
(Eq. [\ref{EQ gamma_R}]) while the dots are the results of a
hydrodynamic simulation.}
\end{figure}

Under the above assumptions and for $\gamma_2 = \gamma_3 \gg 1$ we obtain
\\
\begin{equation}\label{EQ_psi}
\left (\frac{\gamma_4}{\gamma_3}\right )^2 \equiv \psi^2 =
\frac{3a-4}{\sqrt{\frac{12}{a}}(a-1)\,-1}\ ,
\end{equation}
\\
In the limit of a relativistic reverse shock ($a \gg 1$) Eq.
(\ref{EQ_psi}) reduces to $\psi = \gamma_4/\gamma_3\approx
(3a/4)^{1/4}$.

The Lorentz factor, $\gamma$, of the fluid behind the forward shock
as a function of\,\footnote{For convenience we work throughout the
paper in dimensionless variables. Unless specified otherwise,
$\tilde{x} \equiv x/x(R_0)$.}  $\tilde{R}\equiv R/R_0$ is
$\gamma(\tilde{R}<1)=\gamma_4\tilde{R}^{(3-k_0)/2}$ before the
density jump and $\gamma=\psi^{-1}\gamma_4$ immediately after the
density jump. A simple and useful analytic model for
$\gamma(\tilde{R}>1)$ is obtained using the energy conservation
equation while replacing the mass collected up to $R_0$, $M_0=4\pi
A_0R_0^{3-k_0}/(3-k_0)$, by an effective mass $M_{{\rm
eff},0}=\psi^2M_0$. The reasoning behind the factor of $\psi^2$ is
to account for the fact that just after the density jump the bulk
Lorentz factor and the average particle random Lorentz factor in
region 4 ($\gamma_4$) is a factor of $\psi$ higher than that in
region 2 ($\gamma_2 = \gamma_3$), so that the energy in region 4 is
$\approx M_0\gamma_4^2 = \psi^2M_0\gamma^2$. As a result the
expression for energy conservation at $R>R_0$ is approximated by \\
\begin{equation}\label{E_cons}
E = \left[C_0\psi^2M_0+C_1M_1(R)\right]\gamma^2(R)c^2 = {\rm const}\ ,
\end{equation}
\\
where $C_i = 4(3-k_i)/(17-4k_i)$
(this is valid for $k_i<3$), and
\\
\begin{equation}
M_1 = \int_{R_0}^R
4\pi r^2dr\rho_{\rm ext}(r) = M_0\left(\frac{3-k_0}{3-k_1}\right)a
\left(\tilde{R}^{3-k_1}-1\right)\ .
\end{equation}
\\
According to our simple model,
\\
\begin{equation}\label{EQ gamma_R}
\gamma(\tilde{R}>1)=\gamma_4
\left[\psi^2+\left(\frac{17-4k_0}{17-4k_1}\right)a
\left(\tilde{R}^{3-k_1}-1\right)\right]^{-1/2}\ .
\end{equation}
\\ Figure \ref{FIG R_Gamma} shows a comparison of our simple analytic
expression with the results of a hydrodynamic simulation (see appendix
\ref{sec:code} for the simulation details). It depicts the forward
shock Lorentz factor as a function of radius for a spherical
ultra-relativistic blast-wave that propagates into three different
density profiles of the external medium that are described by equation
(\ref{EQ rho_ext}). Two of the density profiles are uniform both below
and above $R_0$ ($k_0 = k_1 = 0$) with density jumps of $a = 10$ and
$a = 100$ at $R_0$. The third density profile presents a wind
termination shock, for which $k_0 = 2$, $k_1 = 0$, and $a = 4$.
Figure \ref{FIG R_Gamma} demonstrates that despite the simplicity of
our analytic approximation, it provides an excellent description of
the accurate solution.

Once the reverse shock reaches $\chi\gtrsim 2$, it samples most of the
energy in the BM radial profile.  At this stage a good fraction of the
total energy is already in region 2, and a similar energy is in region
3. A rough estimate for the radius, $R_1 = R_0+\Delta R$, at which
this occurs may be obtained by using the conditions in region 2 that
have been calculated above according to the shock jump conditions for
a uniform shell, and checking when most of the energy will be in
region 2.  This occurs after a mass $\approx \psi^2 M_0$ is swept from
$r>R_0$, i.e. when the two terms in Eq. \ref{EQ gamma_R} become
comparable, which corresponds to
$(\tilde{R}^{3-k_1}-1)=a^{-1}\psi^2[(17-4k_1)/(17-4k_0)]\sim
a^{-1/2}$, or
\\
\begin{equation}
\tilde{R}_1 = \left[1+\frac{\psi^2}{a}
\left(\frac{17-4k_1}{17-4k_0}\right)\right]^{1/(3-k_1)}\ .
\end{equation}
\\
For $a\gg 1$ this simplifies to $\Delta R/R_0\approx
a^{-1/2}\sqrt{3}(17-4k_1)/[2(17-4k_0)(3-k_1)] \sim a^{-1/2}\ll
1$. Once $E/\gamma^2c^2 = C_0\psi^2 M_0 + C_1M_1 = M_{\rm eff}$
becomes comparable to the mass that would have been swept up at the
same radius if the outer density profile was valid everywhere, $4\pi
A_1R^{3-k_1}/(3-k_1)$, the dynamics approach the new BM self similar
solution for $(A,k)=(A_1,k_1)$. By this time $M_{\rm eff}(R)$ is
dominated by the second term in Eq. \ref{E_cons}, and therefore the
new BM solution is approached when $\tilde{R}^{3-k_1}-1$ becomes
comparable to $\tilde{R}^{3-k_1}$, i.e. when $\tilde{R}\gtrsim
\tilde{R}_{\rm BM} = 2^{1/(3-k_1)}$.

\subsection{The Resulting Light Curve}
\label{sec:rad}

Here the simplified description of the hydrodynamics presented above
is used in order to obtain a semi-analytic expression for the
resulting light curve. We obtain explicit expressions for the three
most relevant power-law segments (PLSs) of the synchrotron spectrum:
$\nu<\nu_m<\nu_c$, $\nu_m<\nu<\nu_c$, and $\nu>\max(\nu_m,\nu_c)$,
where $\nu_m$ is the typical synchrotron frequency and $\nu_c$ is the
cooling frequency. The first two PLSs appear in the slow cooling
regime ($\nu_m < \nu_c$) while the last PLS also appears in the fast
cooling regime ($\nu_m > \nu_c$).

Two useful time scales for calculating the observed radiation are
the radial time, $T_r(R) = t - R/c$, and the angular time,
$T_\theta(R) = R/2c\gamma^2$. The radial time is the arrival time of
a photon emitted at the shock front at radius $R$ along the line of
sight (at $\theta = 0$) relative to a photon emitted at $t=0$ at
$R=0$. The angular time is the arrival time of a photon emitted at
the shock front at an angle of $\theta = \gamma^{-1}$ from the line
of sight relative to a photon emitted at the shock front at the same
radius $R$ along the line of sight. For convenience we normalize the
observed time by $T_0 = T_r(R_0) = R_0/4(4-k_0)c\gamma_4^2$,
$\tilde{T} \equiv T/T_0$.  In our simple model the radial and
angular
times are given by \\
\begin{eqnarray}\nonumber
\tilde{T}_r(\tilde{R}>1) &=& 1+
\left[\psi^2-\left(\frac{17-4k_0}{17-4k_1}\right)a\right]
(4-k_0)\left(\tilde{R}-1\right)
\\ \label{T_r+}
 & &+\left(\frac{4-k_0}{4-k_1}\right)
\left(\frac{17-4k_0}{17-4k_1}\right)
a\left(\tilde{R}^{4-k_1}-1\right)\ ,
\\ \nonumber
\\
\tilde{T}_\theta(\tilde{R}>1) &=&
2(4-k_0)\tilde{R}\left[\psi^2+\left(\frac{17-4k_0}{17-4k_1}\right)
a\left(\tilde{R}^{3-k_1}-1\right)\right]\ ,\quad
\end{eqnarray}
\\
while
\\
\begin{equation}
\tilde{T}_r(\tilde{R}<1) = \tilde{R}^{4-k_0}\ ,\quad
\tilde{T}_\theta(\tilde{R}<1) = 2(4-k_0)\tilde{R}^{4-k_0}\ .
\end{equation}
\\

Following \citet{NP03}, we express the observed flux as an integral
over the radius $R$ of the forward shock. It is convenient to
express the integrand, which represents the contribution from a
given radius $R$ to the observed flux at a given observed time $T$,
as the product of two terms\footnote{This separation is accurate
during the self-similar phase \citep[see][]{NP03} and serves here as
a useful approximation.}: the total emissivity (per unit frequency)
of the blast-wave between $R$ and $R+dR$, $A_\nu(R)$, and a weight
function, $g(\tau,\beta)$, where $\beta = d\log F_\nu/d\log\nu$ is
the spectral slope, which takes into account the relative
contribution from a given radius $R$ to the observed flux at a given
observed time $T$:
\\
\begin{equation}
\tilde{F}_\nu(\tilde{T})  =
C(\beta)\int_0^{\tilde{R}_{max}(\tilde{T)}}
d\tilde{R}\,\,\tilde{A}_{\nu}\,g(\tau,\beta)\ .
\end{equation}
\\
For convenience all variables are normalized by their value at $T_0$
or $R_0$: $\tilde{F}_\nu(\tilde{T}) \equiv
F_\nu(T=\tilde{T}T_0)/F_\nu(T_0)$ and $\tilde{A}_{\nu}(\tilde{R})
\equiv A_\nu(R=\tilde{R}R_0)/A_\nu(R_0)$. The normalization constant
$C(\beta)$ may be obtained by the requirement that
$\tilde{F}_\nu(\tilde{T}=1) = 1$, while $\tilde{R}_{\rm
max}(\tilde{T})$ is given by $T_r[\tilde{R}_{\rm max}(\tilde{T})] =
\tilde{T}$ and may be obtained by inverting equation \ref{T_r+} for
$\tilde{R}$.  The weight function $g(\tau,\beta)$ depends on the the
dimensionless ``time'' variable
\\
\begin{equation}
\tau(\tilde{R},\tilde{T}) \equiv \frac{\tilde{T} -
\tilde{T}_r(\tilde{R})}{\tilde{T}_\theta(\tilde{R})}\ ,
\end{equation}
\\
and on the PLS where the observed frequency $\nu$ is in, which is
specified by the value of the spectral index $\beta$ ($F_\nu \propto
\nu^{\beta}$). In principle, during the self-similar phase, for
$\nu<\nu_c$, $g$ has a complicated form and it also depends on the
power law index, $k$, of the external density \citep{NP03}. However,
the whole approach that is adopted here (of separating the integrand
into the product of $A_\nu$ and a relatively simple $g$ where the
time dependence is only through $\tau$) is anyway not strictly valid
during the non-self-similar phase. Therefore, we choose to use the
simplest expression of $g$ at all PLSs, which is the expression
obtained in the thin emitting region approximation (which is
accurate for $\nu>\nu_c$),
\\
\begin{equation}
g(\tau,\beta) = (1+\tau)^{\beta-2}\ .
\end{equation}
\\

An approximation for $\tilde{A}_\nu(\tilde{R})$ is obtained by
considering three different phases of emission: $\tilde{R} < 1$, $1 <
\tilde{R} < \tilde{R}_1$, and $\tilde{R} > \tilde{R}_1$ (similar to
the approach taken by \citealt{PW06}). For $\tilde{R}<1$ the
blast-wave is self-similar and $\tilde{A}_\nu(\tilde{R}<1)$ is similar
to the one calculated in \citet{NP03}:
\\
\begin{equation}
\tilde{A}_{\nu}(\tilde{R}<1) = \left\{\begin{matrix}
\tilde{R}^{1-k_0} & \quad \nu < \nu_m < \nu_c\ , \cr\cr
\tilde{R}^{2-3p+k_0(3p-5)/4} &
  \quad \nu_m < \nu < \nu_c\ , \cr\cr
\tilde{R}^{1-3p+k_0(3p-2)/4} & \quad \nu >
\max(\nu_m,\nu_c)\ .\end{matrix}\right.
\end{equation}
\\

For $1<\tilde{R}<\tilde{R}_1$ the reverse shock is crossing the hot BM
shell and $\tilde{A}_\nu$ is approximated by evaluating the
contribution from the three emitting regions: $\tilde{A}_{\nu,2}$ --
the shocked external medium originating from $R > R_0$,
$\tilde{A}_{\nu,3}$ -- the portion of the BM shell that has been
shocked by the reverse shock, and $\tilde{A}_{\nu,4}$ -- the
unperturbed portion of the BM solution which has not yet been crossed
by the reverse shock.

Regions 2 and 3 can be approximated as being uniform with the
Lorentz factor given by equation (\ref{EQ gamma_R}). For a
relativistic reverse shock, which occurs for a large density bump
($a\gg 10$), we have $\psi\approx(3a/4)^{1/4}\sim a^{1/4}$. The
ratio of the density behind the forward shock, $n_2\approx 4\gamma_3
n_1$, to that behind the reverse shock, $n_3\approx
4(\gamma_4/2\gamma_3)4\gamma_4 n_0=8(\gamma_4^2/\gamma_3)n_0$, is
$n_2/n_3 \approx a/2\psi^2 \approx \sqrt{a/3}\sim a^{1/2}$.  Since
when both shocks are relativistic the velocity of both shocks
relative to the contact discontinuity is the same, $c/3$, the width
and the volume of regions 2 and 3 is the same (in the contact
discontinuity rest frame), so that their (proper) density ratio is
also the ratio of the rest mass and the total number of electrons
$N_e$ in the two shocked regions. Since $\nu_m\propto\gamma
B\gamma_m^2$ and $\gamma_m\propto e/n$ where $e_2=e_3$ we have
$\gamma_{m2}/\gamma_{m3} = n_3/n_2 \approx \sqrt{3/a} \sim a^{-1/2}$
and $\nu_{m2}/\nu_{m3}\approx 3/a\sim a^{-1}$, where $B \propto
\epsilon_B^{1/2}e^{1/2}$ is the magnetic field measured in the fluid
rest frame and it is assumed that the fraction of the internal
energy in the magnetic field, $\epsilon_B = B^2/8\pi e$, is the same
everywhere.  More generally, for any value of $a>1$, we have
\\
\begin{equation}
\left(\frac{n_3}{n_4}\right)^2 =
\frac{a(3a+\psi^2)}{\psi^2(a+3\psi^2)}\ ,\quad
\frac{n_2}{n_4} = \frac{a}{\psi}\ ,
\end{equation}
\\
and therefore
\\
\begin{equation}
\frac{\nu_{m3}}{\nu_{m2}} = \left(\frac{n_2}{n_3}\right)^2 =
\frac{a(a+3\psi^2)}{3a+\psi^2} \equiv \zeta\ .
\end{equation}
\\ Since $F_{\rm\nu,max}\propto\gamma B N_e$ we have
$F_{\rm\nu,max3}/F_{\rm\nu,max2} = n_3/n_2 = \zeta^{-1/2} \approx
\sqrt{3/a} \sim a^{-1/2}$. Therefore, for $\nu < \nu_m < \nu_c$ or
$\nu_m < \nu < \nu_c$, $F_\nu \propto F_{\rm\nu,max}\nu_m^{\hat{m}}$
and \\
\begin{equation}
\frac{\tilde{A}_{\nu,3}}{\tilde{A}_{\nu,2}}(1<\tilde{R}<\tilde{R}_1)
= \left(\frac{n_2}{n_3}\right)^{2\hat{m}-1} =
\zeta^{\hat{m}-1/2}\ .
\end{equation}
\\
For $\nu > \max(\nu_m,\nu_c)$ the ratio of the spectral emissivity
between regions $2$ and $3$ is equal to the ratio of the energy flux
through the forward and reverse shocks, respectively, that goes into
electrons with a synchrotron frequency close to the observed
frequency, $\nu_{\rm syn}(\gamma_e) \sim \nu$. Since the magnetic
field in the two regions is similar and so is the total energy
flux\footnote{The energy flux is equal in the limit of a
relativistic reverse shock, where the velocities of both shocks
relative regions 2 and 3 is the same ($c/3$). However, even in the
limit of a Newtonian reverse shock ($a-1 \ll 1$) the velocity of the
reverse shock relative to region 3 approaches $c/\sqrt{3}$ (the
sound speed) which is only a factor of $\sqrt{3}$ larger than the
velocity of the forward shock relative to region 2 ($c/3$).}, the
same electron Lorentz factor $\gamma_e$ is required for $\nu_{\rm
syn}(\gamma_e) \sim \nu$, and
\\
\begin{equation}
\frac{\tilde{A}_{\nu,3}}{\tilde{A}_{\nu,2}}(1<\tilde{R}<\tilde{R}_1)
= \left(\frac{\gamma_{m,3}}{\gamma_{m,2}}\right)^{p-2} =
\zeta^{(p-2)/2}\quad\ \ \ \nu > \max(\nu_m,\nu_c)\ .\quad
\end{equation}
\\
Region 4 contributes only to $\nu<\nu_c$. The conditions in this
region still follow the BM solution, since it does not yet know about
the density jump, but the fraction of the BM profile that has still
not passed through the reverse shock decreases with time. This
fraction, $f$, is $1$ at $\tilde{R}=1$ and $\sim 0$ at
$\tilde{R}_1$. We parameterize $f(\tilde{R})$ using a linear
transition with radius,
\\
\begin{equation}
f(\tilde{R}) = \frac{\tilde{R}_1-\tilde{R}}{\tilde{R}_1-1}\ .
\end{equation}
\\
Summing the contributions from all the different regions and
evaluating the contribution from region 2 as $F_\nu \propto
F_{\nu,{\rm max}} \nu_m^{\hat{m}} \propto
M^{\hat{M}}\gamma^{\hat{\gamma}}\rho_{\rm
ext}^{\hat{\rho}}R^{\hat{r}}$ we obtain:
\\
\begin{eqnarray}\nonumber
& \tilde{A}_{\nu} (1<\tilde{R}<\tilde{R}_1)g(\tau,\beta) =
\Theta(\nu_c-\nu)f(\tilde{R}) \tilde{A}_{\nu,\tilde{R}<1}(\tilde{R})
g_{\tilde{R}<1}(\tau,\beta)
\\ \nonumber
& \quad\quad\quad\ +\,
a^{\hat{\rho}}\tilde{R}^{\hat{r}-1-k_1\hat{\rho}}
\left(1+\zeta^{\hat{m}-1/2}\right)
\left[\frac{3-k_0}{3-k_1}\,a\left(\tilde{R}^{3-k_1}-1\right)
\right]^{\hat{M}}
\\  \label{A_nu_R1}
& \quad\quad\quad \times \left[\psi^2+\frac{17-4k_0}{17-4k_1}\,
a\left(\tilde{R}^{3-k_1}-1\right)\right]^{-\tilde{\gamma}/2}
g_{\tilde{R}>1}(\tau,\beta)
\ ,
\end{eqnarray}
\\
where $\Theta(x)$ is the Heaviside step function, and the values of
the exponents for the relevant PLSs are given in Table
\ref{Table:coeff}. The subscript $\tilde{R} < 1$ or $\tilde{R} > 1$
means that the expressions for these $\tilde{R}$ values should be
used, even if the actual value of $\tilde{R}$ does not fall within
this range.

\begin{table}

 \begin{tabular}{lclclclclclcl}
 \hline
 \hline
PLS & $\hat{M}$ & $\hat{r}$ & $\hat{m}$ & $\hat{\gamma}$ &
$\hat{\rho}$\\
 \hline
$\nu<\nu_m<\nu_c$       & 1 & 0 &   $-1/3$  & $2/3$ &   $1/3$   \\
$\nu_m<\nu<\nu_c$       & 1 & 0 & $(p-1)/2$ & $2p$  & $(p+1)/4$ \\
$\nu>\max(\nu_m,\nu_c)$ & 0 & 2 & $(p-1)/2$ & $2p$  & $(p+2)/4$ \\
 \hline
\end{tabular}

\caption{The values of the exponents in equations (\ref{A_nu_R1})
and (\ref{A_nu_R2}) for different power law segments of the
spectrum.} \label{Table:coeff}

\end{table}

At the third phase, $R > R_1$, the reverse shock has finished crossing
the hot BM shell so that only regions 2 and 3 contribute to the
emission. Region 2 gradually relaxes into a self-similar profile,
while region 3 expands and cools at its tail. The contribution from
region 2 remains the same as in the previous phase (i.e. at $R_0 < R <
R_1$). Region 3 does not contribute at $\nu > \nu_c$, while below
$\nu_c$ its contribution can be approximated by assuming that its
hydrodynamic evolution follows that of a fluid element within the tail
of the BM profile, for which $\nu_m(\chi)/\nu_m(\chi=1) =
\chi^{-(37-5k_1)/6(4-k_1)}$ and the peak spectral emissivity per
electron scales as $P_{\nu,e,{\rm max}}(\chi)/P_{\nu,e,{\rm max}}(\chi=1)
= \chi^{-(29-7k_1)/6(4-k_1)}$.  Since during the self-similar
evolution $\chi \propto R^{4-k_1}$ while the number of emitting
electrons in region 3 is constant ($N_e = M_0/m_p$), we obtain
\\
\begin{eqnarray}\nonumber
& \tilde{A}_{\nu} (\tilde{R}>\tilde{R}_1) =
a^{\hat{\rho}}\tilde{R}^{\hat{r}-1-k_1\hat{\rho}}
\left[\psi^2+\frac{17-4k_0}{17-4k_1}\,
a\left(\tilde{R}^{3-k_1}-1\right)\right]^{-\hat{\gamma}/2}
\quad\quad\ \
\\ \label{A_nu_R2}
& \times
\left\{\frac{3-k_0}{3-k_1}\,a\left[\tilde{R}^{3-k_1}-1
+\left(\tilde{R}_1^{3-k_1}-1\right)\,\zeta^{\hat{m}-1/2}
\left(\frac{\tilde{R}}{\tilde{R}_1}\right)^{\hat{\mu}}\right]\right\}^{\hat{M}}
\ ,\quad
\end{eqnarray}
\\
where
\\
\begin{equation}
\hat{\mu}=-\frac{1}{6}\left[29-7k_1+\hat{m}(37-5k_1)\right]\ ,
\end{equation}
\\
and the power-law indices for the three different PLSs is listed in
Table \ref{Table:coeff}.

\section{Case Studies of Spherically Symmetric Jumps in the External Density}
\label{sec: case study}

In this section we study two spherically symmetric external density
profiles which are of special interest for GRB afterglows: a wind
termination shock, and a density jump in a uniform medium.

\subsection{A Wind Termination Shock}
\label{sec:wind}

\begin{figure}
\includegraphics[width=8.5cm]{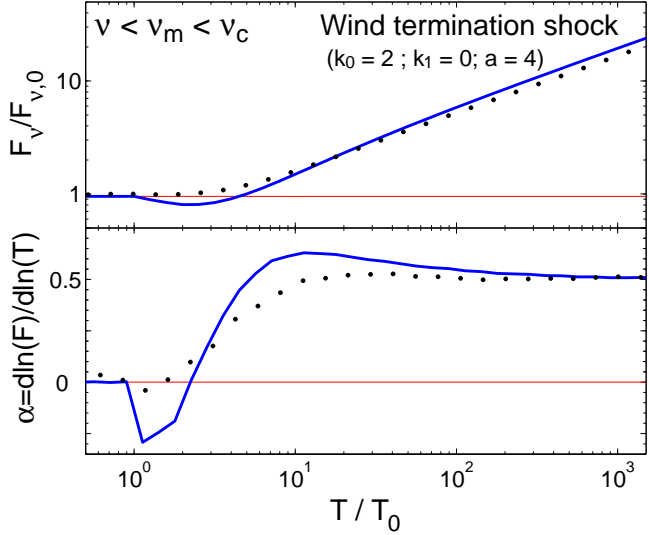}
\caption{\label{FIG wind PLS1}The light curve ({\it upper panel})
and the evolution of the temporal decay index $\alpha$ ({\it lower
panel}) for the synchrotron emission from a spherical relativistic
blast wave running into a wind termination shock ($k_0 = 2$, $k_1 =
0$, $a = 4$ in Eq. [\ref{EQ rho_ext}]), in the frequency range $\nu
< \nu_m < \nu_c$. Shown are the results of the semi-analytic model
described is \S \ref{sec:rad} ({\it solid thick blue line}) and of
the numerical code described in Appendix \ref{sec:code} ({\it black
dots}) for a wind termination shock. The semi-analytic result for
the same wind without a termination shock (where $\rho_{\rm ext} =
A_0 r^{-k_0}$ at all radii; {\it red thin line}) is added for
reference.}
\end{figure}

\begin{figure}
\includegraphics[width=8.5cm]{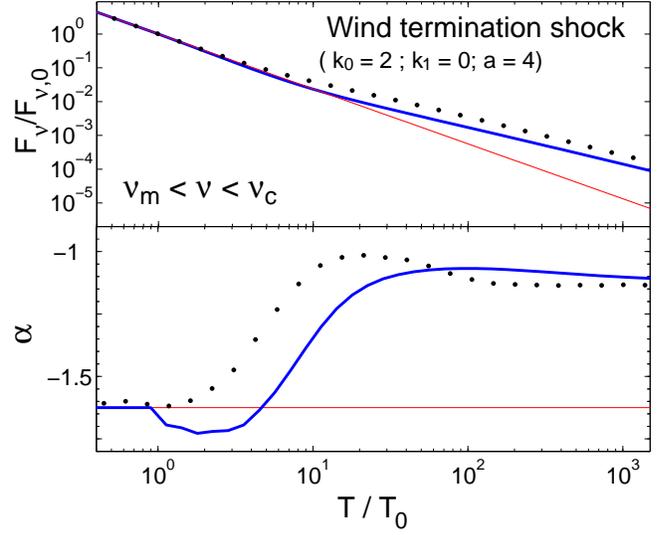}
\caption{\label{FIG wind PLS2}Same as Figure \ref{FIG wind PLS1} but
  for the spectral range $\nu_m < \nu < \nu_c$. We use $p=2.5$.}
\end{figure}

\begin{figure}
\vspace{-0.28cm}
\includegraphics[width=8.5cm]{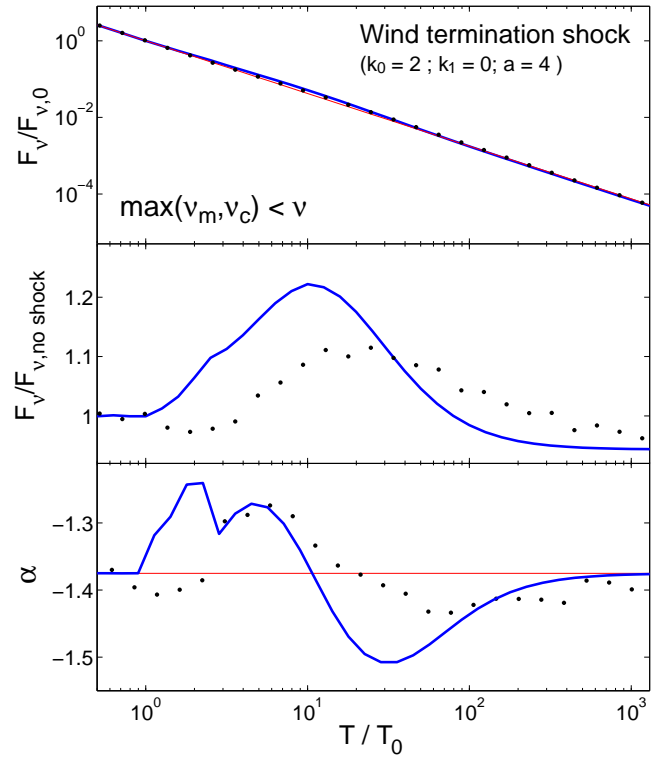}
\caption{\label{FIG wind PLS3} Same as Figure \ref{FIG wind PLS2}
but for the spectral range $\nu > \max(\nu_m,\nu_c)$, and with an
additional panel for the ratio of the flux with and without a wind
termination shock.}
\end{figure}

The semi-analytic model for the light curve that has been developed
in \S \ref{sec:rad} is now applied to a wind termination shock, for
which $k_0 = 2$, $k_1 = 0$, and $a = 4$. We also compare the results
of this semi-analytic model to the numerical model that is described
in Appendix \ref{sec:code}\footnote{ This code assumes optically
thin synchrotron emission and does not take into account opacity
related effects such as synchrotron self-absorption or synchrotron
self-Compton.}. The resulting light curves are displayed in Figures
\ref{FIG wind PLS1}-\ref{FIG wind PLS3} for the three most relevant
power law segments (PLSs) of the spectrum. The results of the
semi-analytic model nicely agree with the numerical results. Some
differences do exist but the qualitative behavior (i.e. variation
time scales and amplitudes) is similar, and even the quantitative
differences are not very large. The main differences between the
semi-analytic model and the numerical simulations are a small
initial dip before the rise in the flux for $\nu < \nu_c$, a
difference in the exact starting time for the change in the temporal
decay index for $\nu_m < \nu < \nu_c$ and $\nu > \max(\nu_m,\nu_c)$,
and a slightly different normalization of the asymptotic flux at $T
\gg T_0$ for $\nu < \nu_c$. The latter arises since we neglect the
dependence of the function $g$ on $k$ in these PLSs ($g$ does not
depend on $k$ for $\nu > \nu_c$) in the semi-analytic model. This
causes a deviation (by a factor of the order of unity) in the
normalization of the asymptotic flux calculated by the semi-analytic
model, compared to its true value, in cases where $k_0 \neq k_1$.

The light-curves show a smooth transition between the asymptotic
power law behavior at $T < T_0$ and at $T \gg T_0$. There is no
rebrightening in PLSs where the flux decays at $T < T_0$ ($\nu >
\nu_m$), and no sharp feature in the light curve which might serve
as a clear observational signature. The transition between the two
asymptotic curves ($T < T_0$ \& $T \gg T_0$) is continuous with the
temporal decay index rising slowly for $\nu<\nu_c$ and mildly
fluctuating for $\nu>\nu_c$. The values of the temporal index
$\alpha \equiv d\log F_\nu/d\log\nu$ during its rising or
fluctuating phase do not exceed its asymptotic value at $T \gg T_0$
by more than $0.1$, at any time. Therefore, the only observable
signature of a wind termination shock is a continuous break (with
$\Delta \alpha = 0.5$) below $\nu_c$. Above $\nu_c$ there is no
change in the asymptotic value of the temporal index $\alpha$, and
it only slowly fluctuates with a very small amplitude ($\approx
0.1$), which is extremely hard to detect. Note that these results
are applicable for a case where the blast-wave remains relativistic
also after it encounters the termination shock (i.e.  $\gamma_3
\approx 0.72 \gamma_4 \gtrsim 3$).

The break in the light curve (the shallowing of the flux decay for
$\nu_m < \nu < \nu_c$ or the transition from constant to rising flux
for $\nu < \nu_m < \nu_c$) occurs over about one decade in time.
Initially there is very little difference relative to the case where
there is no wind termination shock (and $\rho_{\rm ext} =
A_0r^{-k_0}$ at all radii), or even a small dip at $\nu >
\max(\nu_m,\nu_c)$, while a rise in the {\it relative} flux starts
at $\tilde{T} = T/T_0 \sim 2-4$.  The light curve approaches its
asymptotic late time power law behavior at about $\tilde{T} \sim
10-10^2$. This can be understood as follows.

The contribution to the observed flux from within a given angle
$\theta$ around the line of sight does not change drastically across
the density jump. However, there is a sudden decrease in the Lorentz
factor of the shocked material behind the forward shock, so that the
effective visible region increases from $\theta \lesssim
\gamma_4^{-1}$ to $\theta \lesssim \gamma_3^{-1} = \psi/\gamma_4$.
This is responsible for most of the observable signature, and it
starts affecting the light curve noticeably only when photons emitted
just after $R_0$ from an angle $\theta \gtrsim \gamma_4^{-1}$ arrive
to the observer, namely at $\sim T_\theta(R_{0,-}) = 4T_0$ where
\\
\begin{equation}
T_\theta(R_{0,\pm}) \equiv \lim_{\epsilon\to
0}T_\theta[(1\pm\epsilon)R_0]\ .
\end{equation}
\\
This full angular effect becomes apparent when photons from the same
radius and an angle of $\theta \sim \gamma_3^{-1}$ reach the observer,
at $\sim T_\theta(R_{0,+}) = 2(4-k_0)\psi^2T_0 \approx 8 T_0$. The
radial time is smaller than the angular time, and therefore the radial
effect would only slightly increase the time when the total effect
becomes prominent. The light curve approaches its asymptotic power law
behavior when the dynamics approach the new self similar evolution, at
$\sim \tilde{R}_{\rm BM} = 2^{1/(3-k_1)}$. Since
$[\gamma_4/\gamma(\tilde{R}_{\rm BM})]^2 \sim 4$ and therefore
$\tilde{T}_\theta(\tilde{R}_{\rm BM}) \sim 16$, while
$\tilde{T}_r(\tilde{R}_{\rm BM}) \sim 2.6$, this corresponds to
$\tilde{T} \sim 20$. Obviously, this is a rough estimate, but it
agrees reasonably well with the numerical results.

Our main result for the light curves from a wind termination shock
is that there is no prominent readily detectable signature in the
light curve. This is very different from the results of previous
papers that explored a wind termination shock
\citep{R-R01,R-R05,DL02,EGDM06,PW06} which predicted a clear
observational signature (including optical rebrightening also when
the termination shock is at a sufficiently small radius so that the
blast-wave is still relativistic when it runs into it).  In some of
these works \citep{R-R01,R-R05} the forward shock becomes
non-relativistic after hitting the density jump, which might account
for some differences compared to our results which are valid for the
case when the forward shock remains relativistic after running into
the density jump. In other works \citep{DL02,EGDM06,PW06}, however,
the forward shock is assumed to remain relativistic after
encountering the density jump, similar to our assumption. The main
reason for the discrepancy relative to the latter works is that they
did not consider either the effect of the reverse shock on the
dynamics \citep{EGDM06} or did not properly account for the effect
of different photon arrival times from different angles relative to
the line of sight from any given radius \citep{DL02,PW06}. Both
effects tend to smooth out the resulting variability in the light
curve.

\begin{figure}
\vspace{-0.28cm}
\includegraphics[width=8.5cm]{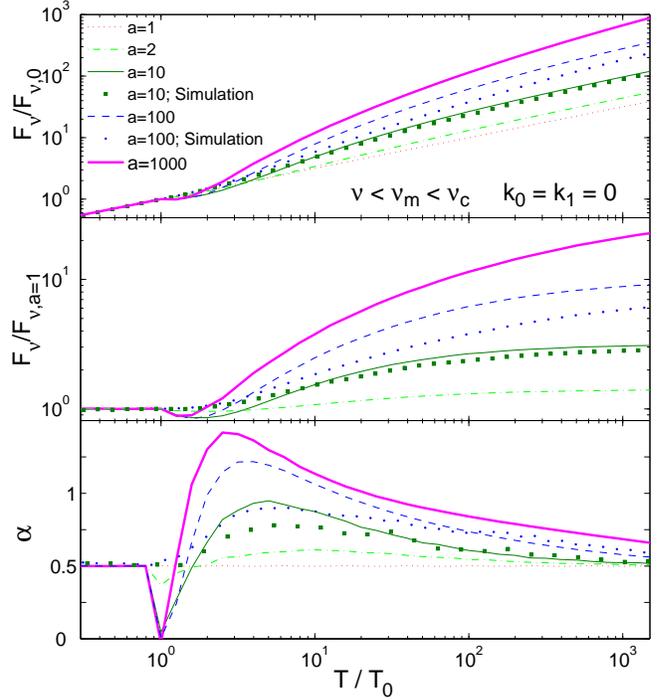}
\caption{\label{FIG ISM PLS1}The light curve ({\it upper panel}), the
ratio of the flux with and without a density jump ({\it middle
panel}), and the evolution of the temporal index $\alpha$ ({\it lower
panel}) for the synchrotron emission from a spherical relativistic
blast wave propagating into a medium with a step like density profile
($k_0 = k_1 = 0$ in Eq.  [\ref{EQ rho_ext}]), in the frequency range
$\nu < \nu_m < \nu_c$.  Shown are the results of the semi-analytic
model described is \S \ref{sec:rad} for four different density
contrasts ($a=2,~10,~100\;\&\;1000$) and of the numerical code
described in Appendix \ref{sec:code} for two cases
($a=10\;\&\;100$). We use here $p=2.5$.}
\end{figure}

\subsection{Spherical Jump in a Uniform External Medium}
\label{sec:uniform}

Next we explore the light curve that results from a spherical
relativistic blast-wave running into a uniform external density with a
jump at some radius $R_0$ (i.e., $k_0 = k_1 =0$ and $a > 1$). Such a
density profile can be generated, for example, by the contact
discontinuity between shocked winds from two evolutionary phases of
the massive star progenitor. This configuration also serves as an
approximation for a large density clump, and constrains the
observational signature from a small density clump (see
\S\ref{sec:clump}).

\begin{figure}
\vspace{-0.30cm}
\includegraphics[width=8.5cm]{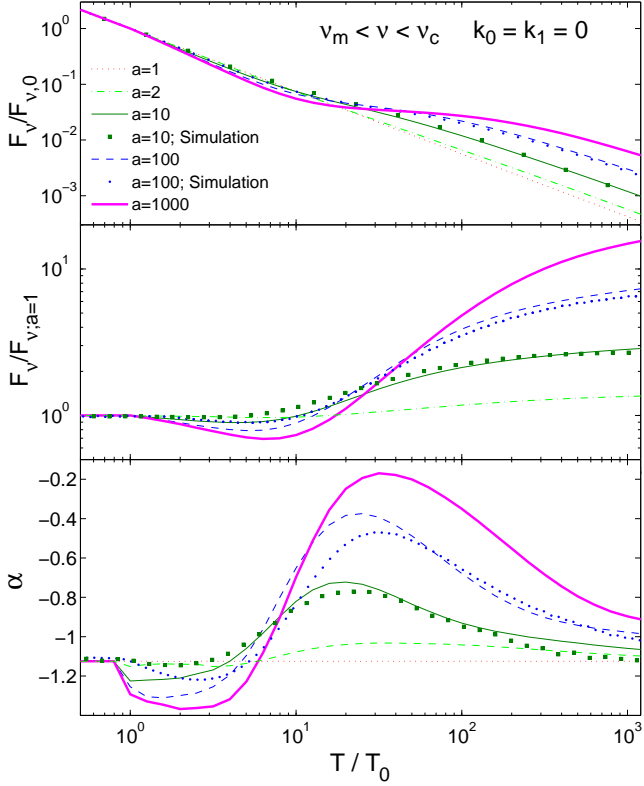}
\caption{\label{FIG ISM PLS2}Same as Figure \ref{FIG ISM PLS1} but
  for the spectral range $\nu_m < \nu < \nu_c$.}
\end{figure}

\begin{figure}
\vspace{-0.30cm}
\includegraphics[width=8.5cm]{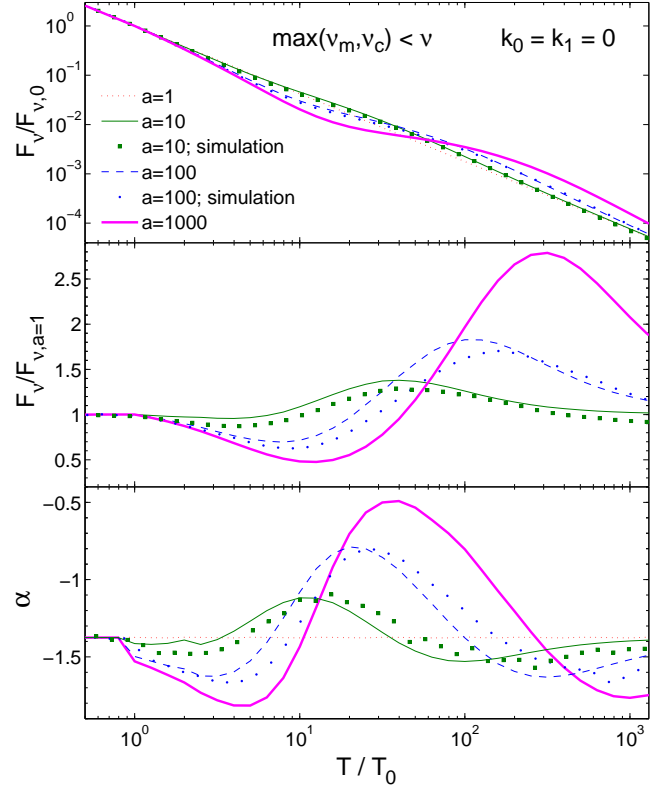}
\caption{\label{FIG ISM PLS3}Same as Figure \ref{FIG ISM PLS1} but
  for the spectral range $\nu_m,\nu_c < \nu$. For clarity we omitted
  the $a=2$ curve.}
\end{figure}

Figures \ref{FIG ISM PLS1}-\ref{FIG ISM PLS3} depict the light
curves from our semi-analytic model (described in \S \ref{sec:rad})
for four different density contrasts ($a=2,\,10,\,100,\,1000$), as
well as the results of the numerical simulation (described in
Appendix \ref{sec:code}) for two of these cases ($a = 10,\,100$).
The agreement between the semi-analytic model and the results of the
simulation is satisfactory. In all cases (all the different PLSs and
$a$ values) the semi-analytic model qualitatively follows the
numerical results and recovers the main features (i.e. the correct
time scales and amplitudes of the variations and their derivatives).
In most cases the quantitative comparison is also impressive (better
than $10\%$). The main differences between the semi-analytic model
and the simulation results are the small initial dip before the rise
in the flux for $\nu < \nu_m < \nu_c$ (observed also in the
wind-termination sock) and an over-shoot for $a = 100$ in this PLS.
The semi-analytic model predicts an initial dip for $\nu_m < \nu <
\nu_c$, which also appears, although less prominently, in the
results of the numerical simulations.

The main result that emerges from Figures \ref{FIG ISM
PLS1}-\ref{FIG ISM PLS3} is that no sharp features appear in any of
the light curves, no matter how high the density contrast (at least
as long as $\gamma_3 = \gamma_4/\psi \gg 1$ where $\psi \approx
(3a/4)^{1/4}$ for $a \gg 1$). Moreover, the maximal deviation of the
temporal decay index, $\alpha$, from its asymptotic value (which is
the same for $T < T_0$ and $T \gg T_0$, since $k_0 = k_1$) is not
large ($< 1$ in all PLSs at all times), and as we show for $\nu_m<
\nu < \nu_c$ it approaches an asymptotic value at large $a$.

\begin{figure}
\includegraphics[width=8.5cm]{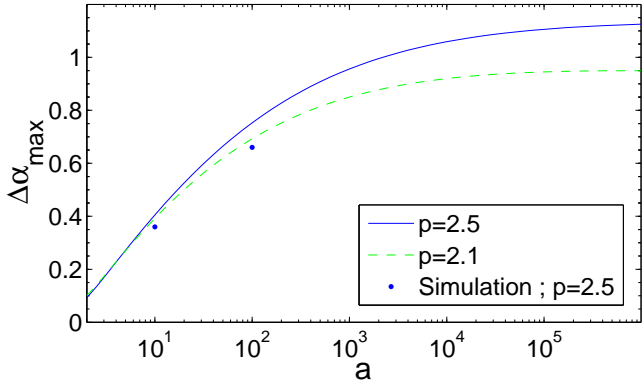}
\caption{\label{FIG Dalpha}The maximal deviation of the temporal
index ($\alpha \equiv d\log F_\nu/d\log T$) from its asymptotic
value, $\Delta \alpha_{\rm max}$, as a function of the density
contrast $a$, for electron power-law indexes $p=2.5$ ({\it solid
line}) and $p=2.1$ ({\it dashed line}). The results of the numerical
simulations for $a=10\;\&\;100$ (in which $p=2.5$) are marked as
dots. }
\end{figure}

Observationally, the most interesting PLS is $\nu_m < \nu < \nu_c$,
since it typically includes the optical. In this PLS the flux
enhancement, $f(T) \equiv F_\nu(T)/ F_\nu(T,a=1)$, is asymptotically
$f(T \gg T_0) = a^{1/2}$, but the transition to this asymptotic value
is very gradual. Figure \ref{FIG Dalpha} depicts the value of the
maximal deviation of the temporal index $\alpha$ from its asymptotic
value, $\Delta \alpha_{\rm max}$, as a function of $a$ for two values
of $p$. We emphasize the behavior of $\Delta \alpha$ since it is
perhaps the easiest quantity to observe. The inability of density
fluctuations to produce sharp features in the light curve is
demonstrated by the low values of $\Delta \alpha_{\rm max}$ that we
find. Some examples are $0.1,~ 0.4~ (0.35),~ 0.75~ (0.65) \;\&\; 0.95$
for $a=2,~10,~100 \;\&\; 1000$, respectively, where the values in
brackets correspond to the results of the numerical
simulation. Furthermore, for very large values of $a$, $\Delta
\alpha_{\rm max}$ saturates at a value of $\approx 1$. Since
$\alpha(T<T_0) = -3(p-1)/4 \sim -1$, no rebrightening (i.e., $\alpha >
0$) is observed. Moreover, at first (just after $T_0$) a mild dip, is
apparent in the light-curve. The depth of this dip increases with
$a$. Another constraining observable is the time over which $\Delta
\alpha_{\rm max}$ is obtained. Thus, we consider the ratio of the time
when $\Delta \alpha_{\rm max}$ is obtained and the time when $\Delta
\alpha > 0$ once it recovers from its initial dip. We find this time
ratio to be $\approx 5$ for any reasonable values of $a$ and $p$.

Our results can be understood as follows. The contribution to the
emission per unit area of the shock front along the line of sight,
from a radius $R$ to an observer time $T$, increases with the energy
density of the shocked fluid and with its bulk Lorentz factor. A
density jump immediately increases the energy density of the freshly
shocked fluid (by a factor $a\,\psi^{-2} > 1$) while reducing its
Lorentz factor (by a factor $\psi^{-1} < 1$). The net effect is such
that the decrement in the Lorentz factor dominates by a small
margin, and the line-of-sight emission actually drops at $R_0$. This
drop increases with $a$ and is the source of the observed dip right
after $T_0$. The same drop in $\gamma$ is also the origin of the
flux increase that follows. With a lower $\gamma$ the largest angle
$\theta$ (from the line of sight) that contributes to the observed
emission increases.  This contribution becomes apparent only when
emission from $\theta \gtrsim 1/\gamma_4$ arrives to the observer,
at $\sim T_{\theta}(R_{0,-})=8T_0$, and is completed when emission
from $\theta \gtrsim 1/\gamma_3$ is observed, at $\sim
T_{\theta}(R_{0,+}) = \psi^2T_\theta(R_{0,-}) \sim a^{1/2}
T_{\theta}(R_{0,-})$. The transition to the asymptotic value
continues up to $T_{\rm BM} \sim a^{1/2} T_{\theta}(R_{0,+}) \sim
10aT_0 $. These time scales explain why the transition is so
gradual, and so is the convergence of $\alpha$ to its asymptotic
value at $T \gg T_0$, for large $a$. Our results show that the
maximal value of $\alpha$ is observed around $T_{\theta}(R_{0,+})$,
and that for $a \gg 1$, $f[T_{\theta}(R_{0,+})] \approx (0.1-0.4)
a^{1/2}$. Now $\Delta \alpha$ can be approximated by
$\log\{f[T_\theta(R_{0,+})]\}/\log[T_{\theta}(R_{0,+})/T_{\theta}(R_{0,-})]$
which approaches $1$ at large $a$.

It is generally accepted that the main signature of density
fluctuations in the external medium are chromatic fluctuations in
the afterglow light curve, where sharp features are expected below
$\nu_c$ (which typically includes the optical bands) and no (or very
weak) variability is expected above $\nu_c$ (which typically
includes the X-rays). This conclusion relies on the fact that a
change in the external density effects the asymptotic light-curve
(at $T \gg T_0$ compared to $T < T_0$) only below $\nu_c$, but not
above $\nu_c$. A comparison between Figures \ref{FIG ISM PLS2} \&
\ref{FIG ISM PLS3} shows that while the behavior in these two PLSs
is indeed different, the general concept described above is
inaccurate and the differences are more subtle. Both PLSs show
smooth fluctuations in the temporal index $\alpha$ with a {\it
comparable} amplitude. The main difference is in the flux
normalization of the asymptotic light curve at $T \gg T_0$ compared
to $T < T_0$.  Above $\nu_c$ the asymptotic light curve does not
change and the observed flux fluctuates around this asymptotic power
law decay, while below $\nu_c$ the normalization for the asymptotic
light curve at $T \gg T_0$ is larger than that at $T < T_0$ by a
factor of $a^{1/2}$, and therefore the flux continuously increase
{\it relative} to the case where there is no density jump ($a = 1$).
This type of difference in the behavior is, however, much harder to
detect (compared to variations in $\alpha$) since, obviously, the
reference light-curve (for $a = 1$) cannot be observed.

Several previous works have explored the light curves arising from
such a density profile \citep{Lazzati02,DL02,NPG03,NP03} and all of
them predicted much sharper features with an observable rebrightening
at $\nu_m<\nu<\nu_c$ (i.e. $\alpha>0$) which does not exist in the
results presented here. In several works \citep{Lazzati02,NPG03,NP03}
the main cause for the discrepancy is that the effect of the reverse
shock on the dynamics was neglected. As we show here, even if the
reverse shock in only mildly relativistic ($a \sim 2$), its effect on
the dynamics cannot be neglected. The reason for this is that even if
the emission from the reverse shock itself is negligible, the
abrupt drop in $\gamma$, which is the Lorentz factor of the shocked
material behind both the reverse and the forward shocks, prevents very
significant variations in $\alpha$ also from the forward shock
emission. \citet{DL02} included a partial consideration of the reverse
shock, however they neglected the strong effect that angular smoothing
has on the light curve.

\subsection{The Effects of Proximity to a Break Frequency}
\label{sec:nu_break}

So far we have assumed that the observed frequency $\nu$ is very far
from the break frequencies ($\nu_m$ and $\nu_c$) and therefore remains
in the same power law segment (PLS) of the synchrotron spectrum
throughout the hydrodynamic transitions that we have
investigated. Under that assumption the flux density normalized by its
value at $T_0$ is independent of frequency within each PLS. In this
subsection we examine the effect on the light curve if the observed
frequency is in the vicinity of a break frequency around the time of
the hydrodynamic transition. For this purpose we use our numerical
code\footnote{Since our semi-analytic model was designed only for the
case where the observed frequency remains in the same PLS, it is not
appropriate for this purpose.} and consider the cases that have been
studied numerically in \S \ref{sec:wind} (a wind termination shock)
and in \S \ref{sec:uniform} (a spherical density jump by a factor of
$a = 10$ or $a = 100$ in a uniform medium).

\begin{figure}
\vspace{-0.28cm}
\includegraphics[width=8.5cm]{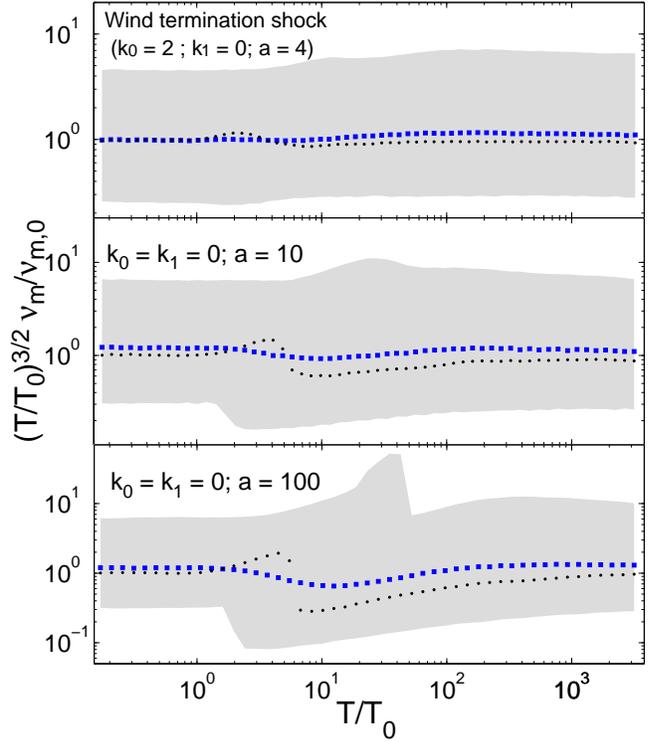}
\caption{\label{FIG:nu_m}The temporal evolution of the typical
  synchrotron frequency $\nu_m$ ({\it dotted line}), which is defined
  as the frequency at which the spectral index $\beta$ is midway
  between its value at $\nu \ll \nu_m$ ($\beta_1 = 1/3$) and at $\nu
  \gg \nu_m$ [$\beta_2 = (1-p)/2$], $\beta = (\beta_1+\beta_2)/2$.  We
  use $p = 2.5$. The shaded region shows the frequency range where
  80\% of the change in $\beta$ occurs (between 10\% and 90\% of
  $\Delta\beta = \beta_1-\beta_2$). The frequency where the
  asymptotic power laws well above and  below $\nu_m$ meet is
  marked with filled squares. For clarity, all frequencies are
  normalized by $\nu_{m,0} = \nu_m(T_0)$ and multiplied by the
  appropriate power of $\tilde{T}= T/T_0$ which takes out the
  asymptotic time dependence of $\nu_m$ at early and late times. The
  different panels are for different spherical density profiles with a
  sharp density jump that have been studied in \S
  \ref{sec:wind} -- a wind termination shock ({\it upper panel}), and
  in \S \ref{sec:uniform} -- a spherical density jump by a factor of
  $a = 10$ ({\it middle panel}) and $a = 100$ ({\it lower panel}) in a
  uniform medium.}
\end{figure}

\begin{figure}
\vspace{-0.28cm}
\includegraphics[width=8.5cm]{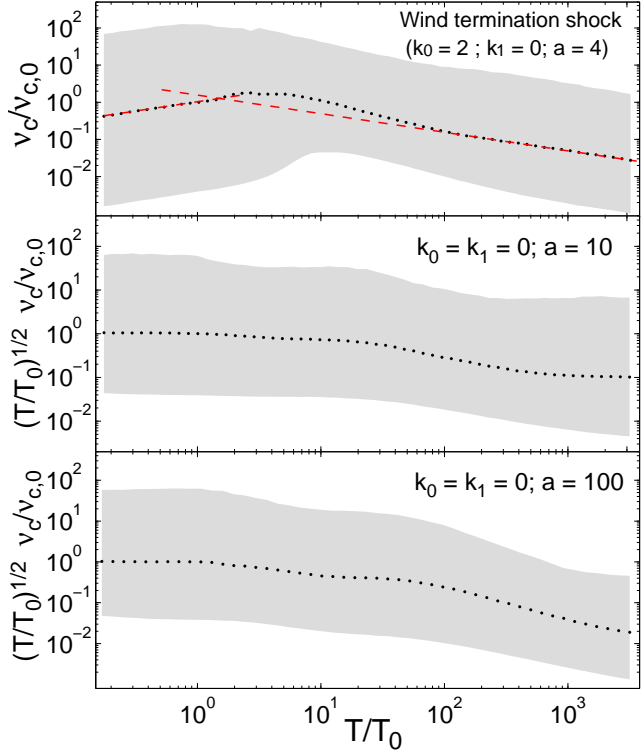}
\caption{\label{FIG:nu_c}The temporal evolution of the cooling break
  frequency, $\nu_c$, in the same format as Figure \ref{FIG:nu_m};
  here $\beta_1 = (1-p)/2$, $\beta_2 = -p/2$, and again $p = 2.5$. For
  a wind termination shock ({\it upper panel}) the asymptotic temporal
  index at $T < T_0$ ($\alpha_1 = 1/2$; {\it left dashed line}) is
  different from that at $T \gg T_0$ ($\alpha_2 = -1/2$; {\it right
  dashed line}), and therefore the normalized frequencies are not
  multiplied by $(T/T_0)^{1/2}$ as in the other two panels (that are
  for a density jump in a uniform medium for which $\alpha_1 =
  \alpha_2 = -1/2$).}
\end{figure}

Figures \ref{FIG:nu_m} and \ref{FIG:nu_c} show the temporal evolution
of the spectral break frequencies $\nu_m$ and $\nu_c$, respectively,
around the time of the density jump. The break frequencies are defined
as\footnote{Defining instead $\nu_b =
\max\{\nu|\beta(\nu)>0.5\beta_1+0.5\beta_2\}$ makes no noticeable
difference.} $\nu_b = \min\{\nu|\beta(\nu)<0.5\beta_1+0.5\beta_2\}$,
where $\beta_1$ and $\beta_2$ are the asymptotic values of the
spectral index $\beta$ at $\nu \ll \nu_b$ and $\nu \gg \nu_b$,
respectively, and $b = m,\,c$. The shaded region shows the frequency
range $\nu_{10\%} < \nu < \nu_{90\%}$ where $\nu_{10\%} =
\min\{\nu|\beta(\nu)<0.9\beta_1+0.1\beta_2\}$ and $\nu_{90\%} =
\max\{\nu|\beta(\nu)>0.1\beta_1+0.9\beta_2\}$, i.e. where 80\% of the
change in $\beta$ across the spectral break occurs.

The typical synchrotron frequency $\nu_m$ fluctuates around its
asymptotic $T^{-3/2}$ power law decay, which does not depend on the
power law index $k$ of the external density. Furthermore, even for a
wind termination shock there is no observable change in the asymptotic
normalization (i.e. the asymptotic value of $T^{3/2}\nu_m$) and only a
very small change in the width of the spectral break which is depicted
by the shaded region \citep[in agreement with the semi-analytic
results of][]{GS02}. For a density jump in a uniform medium there is
no change in the asymptotic normalization or in the asymptotic shape
of the spectral break, again as expected from analytic
calculations. Both the amplitude and the typical time scale of the
fluctuations in $T^{3/2}\nu_m$ increase with the density contrast $a$,
where the amplitude scales roughly as $a^{1/2}$ and the time scale is
roughly linear in $a$.  There is a sharp feature in $\nu_m$ which
occurs first for $\nu_{m,10\%}$, then for $\nu_m$, and finally for
$\nu_{m,90\%}$.  This can be understood as follows. Immediately after
the forward shock encounters the density jump $\nu_m$ decreases in
region 2 and increases in region 3, by a factor of $\sim a^{1/2}$ in
both cases.  Therefore, as long as the region 3 contributes
significantly to the observed spectrum, the break is a superposition
of two peaks (corresponding to $\nu_{m,2}$ and $\nu_{m,3}$), separated
by a factor of $\sim a$ in frequency, and thus its width increases
significantly. Moreover, it causes $\beta(\nu)$ to be non-monotonic,
and the definitions of $\nu_{m,10\%}$, $\nu_m$, and $\nu_{m,90\%}$
cause these frequencies to have a finite jump when the extremum in
$\beta(\nu)$ crosses the appropriate value of $\beta$, which occurs
later for higher frequencies. Given the complex structure of the
spectral break around $\nu_m$ we also present in figure \ref{FIG:nu_m}
the evolution of the frequency where the asymptotic power laws well
above and below the break meet \citep[which can serve as an
alternative definition for the location of the break frequency, as was
done in][]{GS02}. This frequency evolves very smoothly and shows very
mild fluctuations in all cases, since it is less effected by the
transient broadening of the spectral break during the hydrodynamic
transition.

\begin{figure}
\vspace{-0.28cm}
\includegraphics[width=8.5cm]{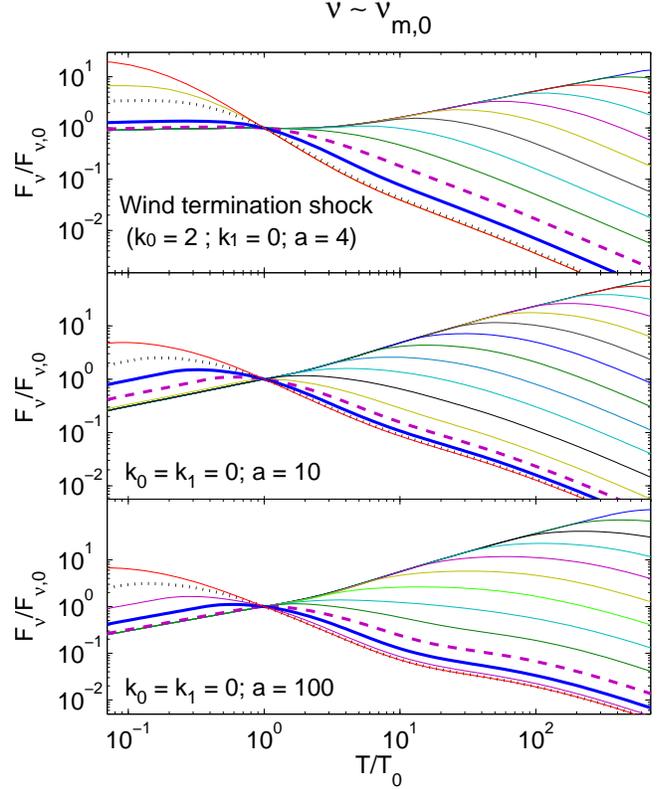}
\caption{\label{FIG:LC,nu_m}The light curves for frequencies that are
  in the vicinity of $\nu_m$ at the time of the density jump, for the
  same hydrodynamic transitions that are shown in Figure
  \ref{FIG:nu_m} (with $p = 2.5$). The {\it solid thick line} is for
  $\nu \approx \nu_m(T_0)$, the {\it dashed thick line} is for $\nu
  \approx \nu_{m,10\%}(T_0)$, and the {\it dotted thick line} is for
  $\nu \approx \nu_{m,90\%}(T_0)$. The remaining curves are spaced by
  a factor of 2.5 in frequency in the middle panel and 3 in the top
  and bottom panels.}
\end{figure}

The evolution of the cooling break frequency, $\nu_c$, is shown in
Figure \ref{FIG:nu_c}. For a wind termination shock the temporal
index $\alpha$ has different asymptotic values at $T<T_0$ ($\alpha_1
= 1/2$) and at $T \gg T_0$ ($\alpha_2 = -1/2$); $\nu_c$ transitions
rather smoothly between these two asymptotic limits, with a slight
overshoot (i.e. $d\log\nu_c/d\log T$ dips below $-1/2$) due to the
increase in density across the jump (the asymptotic value of $\nu_c$
decreases with increasing density). The asymptotic behavior of
$\nu_c$ is marked with dashed lines showing that for practical
purposes (e.g., analytic calculation) it can be well approximated as
a sharp temporal transition between $\alpha1$ and $\alpha_2$ at $T
\approx 1.5 T_0$, while keeping in mind that the spectral break
itself is very smooth at any time.  For a density jump in a uniform
medium the asymptotic value of the temporal index does not change
($\alpha_1 = \alpha_2 = -1/2$), but the asymptotic normalization of
$T^{1/2}\nu_c$ at $T \gg T_0$ is a factor of $a$ lower than at $T <
T_0$. The transition between the two asymptotic limits is fairly
smooth. The shaded region which corresponds to 80\% of the change in
$\beta$ across the break is significantly larger for $\nu_c$
compared to $\nu_m$, corresponding to a smoother break, in agreement
with semi-analytic calculations \citep{GS02}.  The transition to the
late time asymptotic behavior stretches over a larger factor in
time, which increases with $a$.

Figures \ref{FIG:LC,nu_m} and \ref{FIG:LC,nu_c} show light curves for
frequencies that are in the proximity of $\nu_m$ and $\nu_c$,
respectively, around the time of the density jump. There is a smooth
transition between the light curves for frequencies that are well
below the break frequency and those for frequencies well above the
break frequency around the time of the density jump. Our main
conclusions from sections \S\S \ref{sec:wind} and \ref{sec:uniform}
remain valid also when the observed frequency is near a break
frequency around the time of the density jump. In particular, there is
no rebrightening at $\nu \gg \nu_m$, and the observed features in the
light curve are very smooth. Therefore, our main results are rather
robust.

\begin{figure}
\vspace{-0.28cm}
\includegraphics[width=8.5cm]{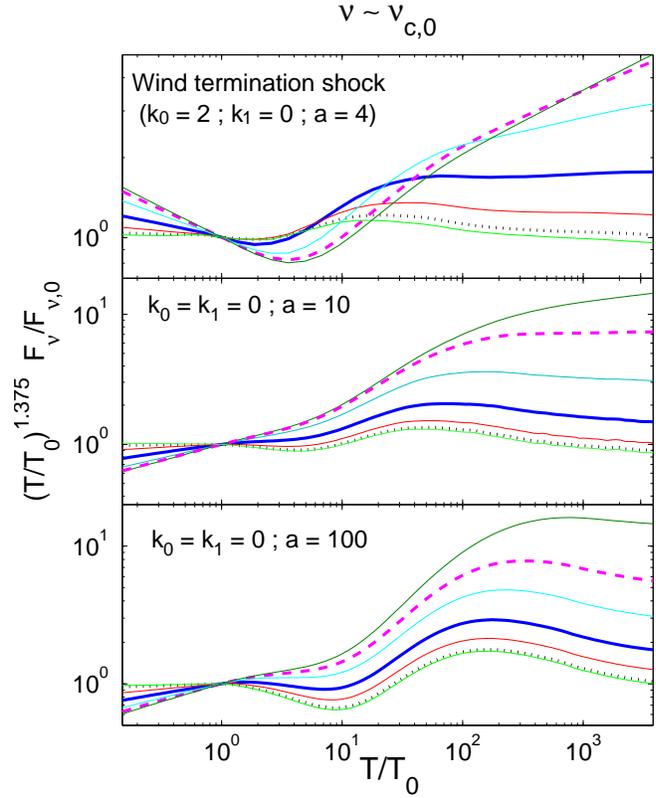}
\caption{\label{FIG:LC,nu_c}The light curves for frequencies near
  $\nu_c$ at the time of the density jump, for the same hydrodynamic
  transitions that are shown in Figure \ref{FIG:nu_c} (with $p =
  2.5$). The normalized flux is multiplied by $(T/T_0)^{(3p-2)/4}\to
  \tilde{T}^{1.375}$ in order to eliminate asymptotic time dependence
  at $T \gg T_0$ for a uniform medium. The {\it solid thick line} is
  for $\nu \approx \nu_c(T_0)$, the {\it dashed thick line} is for
  $\nu \approx \nu_{c,10\%}(T_0)$, and the {\it dotted thick line} is
  for $\nu \approx \nu_{c,90\%}(T_0)$. The remaining curves are spaced
  by a factor of $\approx 15$ in frequency in the top panel and
  $\approx 8$ in the middle and bottom panels.}
\end{figure}

\section{A Clump in the External Density}
\label{sec:clump}

In this section we estimate the effect of a clump in the external
density on the light curve. By a clump we refer to a well localized
region of typical size $l_{\rm cl}$ which is overdense by a factor of
$a > 1$ relative to the uniform background external density. For a
given clump size $l_{\rm cl}$ and overdensity $a$ (well within the
clump, near its center) the effect on the light curve is expected to
be larger if the clump has sharper edges, i.e.  the smaller the length
scale $\Delta l$ over which the density rises by a factor of $\approx
a$ relative to the background density. While in practice one might, in
many cases, expect $\Delta l \sim l_{\rm cl}$, we consider the limit
of a sharp edged clump with $\Delta l \ll l_{\rm cl}$, in order to
maximize the effect on the light curve.

Because of relativistic beaming, most of the contribution to the
observed light curve from a given radius $R$ is from within an angle
of $\theta \lesssim \gamma^{-1}$ around the line of sight, which
corresponds to a lateral size of $\sim R/\gamma$.  Therefore, a clump
in the external density of size\footnote{Here $\gamma$ is the Lorentz
factor inside the clump, which is smaller than that before the
afterglow shock hits the clump, by a factor $\psi$ that is given in
equation (\ref{EQ_psi}).} $l_{\rm cl} \gg R/\gamma$ would not differ
considerably from a spherically symmetric density jump that was
considered in \S \ref{sec:spherical}. If the surface of the clump is
not normal to the line of sight (e.g. if the line of sight to the
central source does not pass through the center of a spherical clump),
this is expected to reduce the effect of the clump on the light curve
(similar to what is expected if the clump does not have very sharp
edges, $\Delta l \sim l_{\rm cl}$ rather than $\ll l_{\rm
cl}$). Therefore the results of \S \ref{sec:spherical} can be viewed
as a rough upper limit on the effect of ``big'' clumps ($l_{\rm cl} >
R_0/\gamma_3$). Small to intermediate clumps, of size $l_{\rm cl}
\lesssim R_0/\gamma_3$ are expected to have a smaller effect on the
light curve and are investigated below.

\begin{figure}
\includegraphics[width=8.5cm]{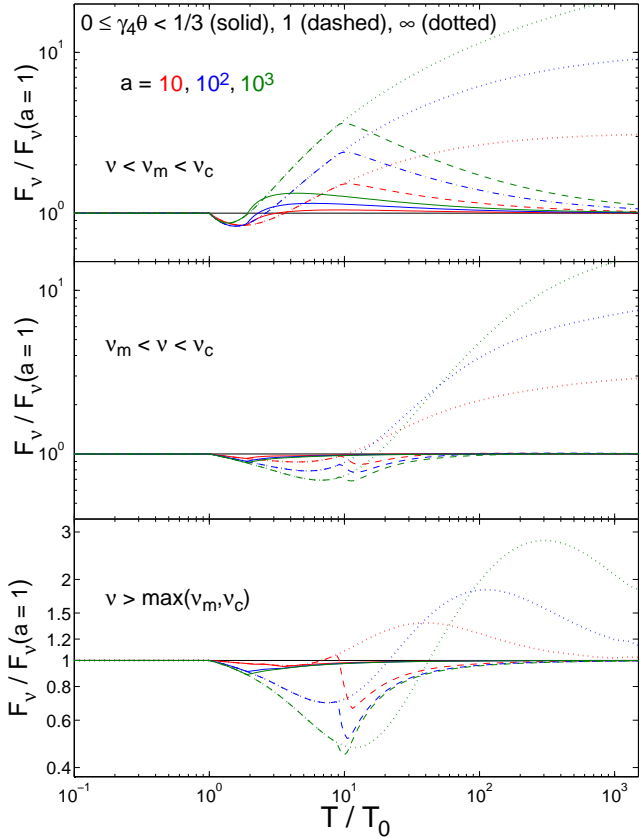}
\caption{\label{FIG clumps}The ratio of the flux with and without a
  density jump at $R > R_0$, within a finite angle $\theta <
  \theta_{\rm max}$ around the line of sight, for three different
  density contrasts ($a = 10,\,100,\,1000$) and three different
  angular sizes ($\gamma_4\theta_{\rm max} = 1/3,\,1,\,\infty$).  This
  serves as an approximate upper limit for the effect of a clump in
  the external medium on the light curve. The different panels are for
  the most relevant power law segments of the synchrotron spectrum.}
\end{figure}

The results of the previous section, and in particular the
semi-analytic model for the light curve that was derived in \S
\ref{sec:rad}, can be used to put an approximate upper limit on the
effect that a density clump could have on the observed light
curve. Such a limit is achieved by using the spherical model from \S
\ref{sec:rad} within a finite solid angle: $\theta_{\rm min} < \theta
< \theta_{\rm max}$ and $\phi_{\rm min} < \phi < \phi_{\rm max}$ in
spherical coordinates, while in the radial direction the clump extends
out to $R \gg R_0$. In practice we expect the radial extent of the
clump to be $\Delta R \sim l_{\rm cl}$, similar to its extent in the
$\theta$ direction ($\sim R_0\Delta\theta$ where $\Delta\theta =
\theta_{\rm max} - \theta_{\rm min}$) and in the $\phi$ direction
($\sim R_0\bar{\theta}\Delta\phi$ where $\Delta\phi = \phi_{\rm max} -
\phi_{\rm min}$ and $\bar{\theta} = [\theta_{\rm min} + \theta_{\rm
max}]/2$). In our coarse approximation the clump has no upper bound in
the radial direction, and its lateral size scales linearly with
radius. Obviously, this sets an (approximate) upper limit for the
effect on the light curve of a clump with a size $l_{\rm cl}$ in all
directions (which is roughly given by $l_{\rm cl} \sim R_0\Delta\theta
\sim R_0\bar{\theta}\Delta\phi \sim \Delta R$ in spherical
coordinates).

\begin{figure}
\includegraphics[width=8.5cm]{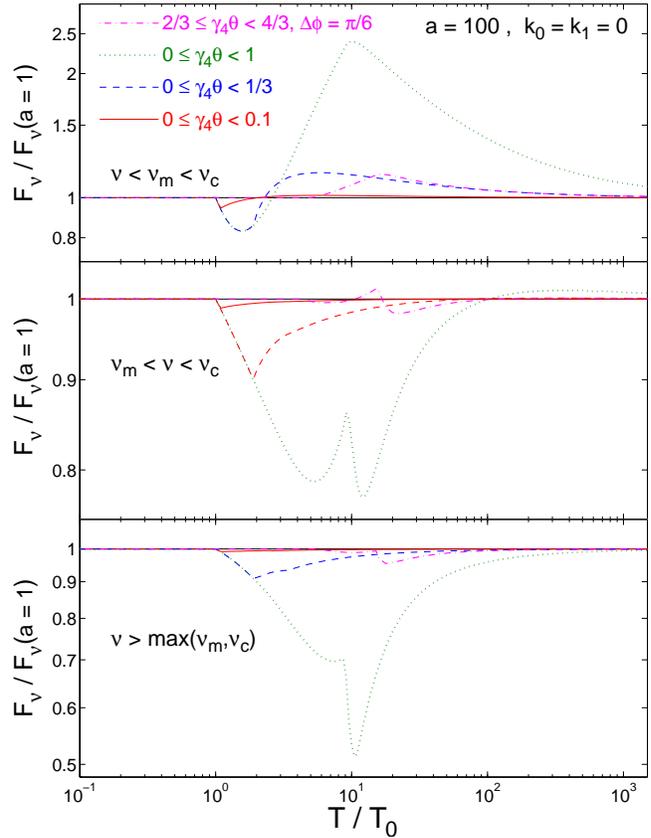}
\caption{\label{FIG clumps2}Similar to Figure \ref{FIG clumps} but for
  a fixed density contrast, $a = 100$, and for three clumps along the
  line of sight together with a clump that is at the side of the
  visible region at the time of collision, centered around $\gamma_4\theta
  \approx 1$.}
\end{figure}

Figure \ref{FIG clumps} shows the results of this model for a
density clump that lies along the line of sight, $\theta_{\rm min} =
0$ and $\gamma_4\theta_{\rm max} = 1/3,\,1,\,\infty$ while
$\Delta\phi = 2\pi$, for three different values of the density
contrast ($a = 10,\,100,\,1000$) and for the three most relevant
power law segments of the spectrum (that were modeled in \S
\ref{sec:rad}). The smaller the angular extent of the clump, the
smaller the amplitude of the change in the flux {\it relative} to
its value for the smooth underlying density distribution without the
clump ($a = 1$), and the smaller the factor in time over which it
effects the flux significantly (e.g., the full width at half maximum
of the {\it relative} flux). This behavior is expected since
both the amplitude and the duration of the fluctuation depends on
the size of the clump \citep{IKZ05}. The amplitude depend on the ratio
in size between the perturbed region (of length scale $l_{cl}$) and
the unperturbed region (of length scale $R/\gamma$) of the blast
wave, and thus increase with $l_{cl}$. The duration depends on the
delay in the arrival time of photons emitted from the perturbed
region, which again increase with the size of this region.

As can be seen in Figure \ref{FIG clumps}, the amplitude of the
fluctuation in the relative flux increases with the density contrast
$a$. There is a sharp transition in the light curve at $\tilde{T} =
1 + 2(4-k_0)(\gamma_4\theta_{\rm max})^2$, when the radiation from
the outer edge of the clump at $R = R_0$ reaches the observer, which
corresponds to the peak in the relative flux for
$\gamma_3\theta_{\rm max} = \gamma_4\theta_{\rm max}\psi^{-1} \sim
1$.  Such sharp features (see also Figure \ref{FIG clumps2}) are
caused by the over-simplified clump model that we use here, and are
expected to be smoothed out in more realistic models of clumps.
Above $\nu_m$ a clump can produce a dip or a bump in the relative
flux with a relatively small amplitude (depending on its size and
density contrast), while below $\nu_m$ it produces a bump in the
relative flux with a larger amplitude.

Figure \ref{FIG clumps2} shows the {\it relative} flux for a fixed
density contrast, $a = 100$, for three clumps along the line of sight
with different angular sizes ($\Delta\phi = 2\pi$, $\theta_{\rm min} =
0$ and $\gamma_4\theta_{\rm max} = 0.1,\,1/3,\,1$), as well as for a
clump close to the edge of the visible region at the time of the
collision, $\gamma_4(\theta_{\rm min},\theta_{\rm max}) = (2/3,4/3)$
and $\Delta\phi = \pi/6$, which occupies the same solid angle as the
clump along the line of sight with $\gamma_4\theta_{\rm max} = 1/3$.
The effect of a given clump on the light curve is maximal when it is
along the line of sight, and it becomes smaller the further it is from
the line of sight (i.e., its effect is significant over a shorter time
scale and the amplitude of the change in the relative flux is somewhat
reduced).

Overall, a fairly large clump of size $l_{\rm cl} \gtrsim R/\gamma$
with a sufficiently large density contrast ($a \gtrsim 10-10^2$) is
required in order to produce an observable signature in the light
curve, and even then the most prominent signal would be below
$\nu_m$, which is typically relevant for the radio. In the optical,
which is typically above $\nu_m$, there would only be a small dip or
bump in the relative flux, which would be hard to detect.

\section{Implications to GRBs 021004, 000301C, 030329 and SHB
060313}\label{sec:GRBs}

Next we reconsider the cause for the fluctuations in the optical
afterglow light curves of the long-soft GRBs 021004, 000301C, and
030329, as well as the very recent short-hard GRB 060313, in light
of our results. GRB~021004 showed significant variability in its
optical afterglow \citep{Pandey02,Fox03,Bersier03,Uemura03} which
included three distinct episodes of mild rebrightening ($\alpha
\gtrsim 0$) at $T \sim 0.05\;$days, $\sim 0.8\;$days, and $\sim
2.6\;$days. Just before the first of these epochs, $\alpha \sim
-0.7$, and it then became positive over a factor of $\sim 2-3$ in
time. Such a large increase in $\alpha$ over a relatively small
factor in time is hard to accommodate by a jump in the external
density: $\Delta \alpha \sim 1$ requires $a \gtrsim 10^2$ and even
then it is hard to achieve such an increase in $\alpha$ over a
factor of $\sim 2-3$ in time. The second rebrightening episode lies
within the tail of the first, and is therefore not a very ``clean''
case to study. The third rebrightening epoch at $\sim 2.6\;$days is
somewhat more isolated, and has $\Delta\alpha \gtrsim 1$ where most
of the increase in $\alpha$ is within about a factor of $\sim 1.5$
in time. This is extremely hard to achieve by variations in the
external density. On the other hand, angular fluctuations in the
energy per solid angle within the jet (a ``patchy shell'') nicely
account for both the fluctuations in the light curve and the
variability in the linear polarization of this afterglow
\citep{GK03,NO04}.

GRB~000301C displayed a largely achromatic bump in its optical to
NIR light curves \citep{Sagar00,Berger00} peaking at $T \sim
3.8\;$days. Before the bump $\alpha \approx -(1.2-1.3)$ and during
the rise to the bump $\alpha$ became slightly positive, so that
$\Delta\alpha \sim 1.5$, where most of the increase in $\alpha$
occurred over a factor of $\sim 1.5$ in time. We find that this
cannot be produced by a sudden change in the external density
\citep[as has been suggested by][]{Berger00}.  The decay just after
the peak of the bump is very sharp, $\alpha \approx -(3.5-4)$, and
since $\beta -\alpha > 2$ \citep[][find $\beta = -0.96\pm 0.08$
during the decay just after the peak, at $T = 4.8\;$days]{Sagar00}
this requires a deviation from spherical symmetry \citep{KP00}, and
might not be easy to achieve with angular fluctuations in the energy
per solid angle within the jet (a ``patchy shell'') or aspherical
refreshed shocks. In this case an alternative microlensing
interpretation \citep{GLS00} was shown to be able to nicely
reproduce the shape of the bump \citep{GGL01}.

GRB~030329 has one of the best monitored and densely sampled optical
afterglow light curves to date. It is presented in great detail by
\cite{Lipkin04} who clearly show that the light curve contains
several rebrightening ($\alpha>0$) episodes in which $\Delta \alpha>
2$. All of these episodes have a similar structure and occur over a
small factor in time ($\Delta T <T$). In previous works it has been
suggested that some of these rebrightening episodes are a result of
density fluctuations. \citet{PW06} have suggested that the first
(and largest) rebrightening episode is the signature of a wind
termination shock. \citet{Sheth03} have followed \citet{Berger03} in
attributing the first rebrightening episode to a two-component jet,
but argued that the subsequent rebrightening episodes could be
explained by variations in the external density. Our results clearly
show that none of the bumps in the optical light curve of GRB~030329
can be a result of density bumps. This implies that the two-jet
model of \citet{Berger03} and \citet{Sheth03} fails to account for
most of the observed light-curve fluctuations. Excluding density
fluctuations as the possible source of any of the major bumps,
together with the similarity in the shapes of the bumps
\citep{Lipkin04}, support a sequence of similar episodes in which
the energy of the jet fluctuates, such as a late time energy
injection by ``refreshed shocks'' \citep{GNP03}.

The afterglow of the very recent short-hard GRB~060313 has been
monitored both by the X-ray telescope (XRT) and by the
optical/ultra-violate telescope (UVOT) on board {\it Swift}
\citep{Roming06}. The optical/UV light curve showed three sharp
bumps/flares at $T \sim 1.7\;$hr, $\sim 3.2\;$hr, and $\sim
6.6\;$hr, with an amplitude of more than a factor of 2 in flux and a
very short rise time of $\Delta T \lesssim 0.1T$. During the same
time the X-ray light curve showed a smooth (and steeper) power law
decay. This was interpreted by \citet{Roming06} as the result of
variations in the external density by a factor of $\sim 2$, where
the lack of variability in the X-rays was attributed to $\nu_c$
being between the optical/UV and the X-rays. Our results clearly
show that such sharp rebrightening episodes as were seen in the
optical/UV afterglow light curve of GRB~060313 cannot be the result
of variations in the external density.  Therefore, there must be
some other cause for this variability, such as late time internal
shocks with a very soft spectrum, as was suggested by
\citet{Roming06} as an alternative mechanism.

\section{Conclusions}
\label{sec:dis}

We have presented a semi-analytic model for the light-curve resulting
from synchrotron emission by a spherical relativistic blast-wave that
propagates into a power law external density profile with a single
sharp density jump at some radius $R_0$. Our solution is general
enough to include a transition in the density power-law index ($k$) at
$R_0$, but is limited to cases in which the blast-wave remains
relativistic after it encounters the density jump. This model has been
used to explore in detail two density profiles that are most relevant
to GRB afterglows: a wind termination shock, and a sharp density jump
between two regions of uniform density. The latter results are also
used to constrain the signature of density clumps in a uniform
medium. We have also carried out detailed numerical simulations for
several of the cases which we have studied in detail. These numerical
results serve three purposes. First, they are used in order to obtain
more accurate light curves for the cases which are of special
interest.  Second, they are appropriate for calculating the light
curves in the vicinity of a spectral break frequency (\S
\ref{sec:nu_break}).  Third, they serve as a test for the quality of
our simple semi-analytic model, which is found to give a very good
qualitative description and reasonable quantitative description of the
light curve.

Our main result is that density jumps do not produce sharp features in
the light curve, regardless of their density contrast! The results of
our specific case studies are as follows.\vspace{0.1 in}\linebreak A wind
termination shock:
\begin{itemize}
    \item The light curve shows a smooth transition,
    which lasts for about one decade in time, between the asymptotic
    power-law behavior at $T < T_0$ and at $T \gg T_0$.

    \item There is no rebrightening or any other sharp feature that
    can be used as a clear observational signature.

    \item Above the cooling frequency, $\nu_c$, there is no change in
    the asymptotic value of the temporal index, $\alpha$, and it only
    fluctuates with a small amplitude ($\Delta \alpha \approx 0.1$).

    \item The only observable signatures of a wind termination shock
    are a smooth break, with an increase of $\Delta \alpha = 0.5$ in
    $\alpha$, below $\nu_c$ and a transition in the temporal evolution
    of $\nu_c$.

\end{itemize}
A density jump between two uniform density regions:
\begin{itemize}
    \item The light curve shows a smooth transition between
    the two asymptotic power laws (at $T < T_0$ and $T \gg T_0$).

    \item The transition time increases with the density contrast, $a$,
    and is about $\sim 10aT_0$.

    \item The maximal deviation, $\Delta \alpha_{\rm max}$, of the
    temporal index $\alpha$ from its asymptotic value (at $T < T_0$
    and $T \gg T_0$) is small. For example, $\Delta \alpha_{\rm
    max}(a=10) < 0.4$; $\Delta \alpha_{\rm max}$ depends weakly on $a$
    and approaches $\approx 1$ at very large $a$ values.  Therefore,
    a density jump cannot produce an optical rebrightening when
    $\nu_{\rm optical} > \nu_m$.

    \item The light curve fluctuates also above $\nu_c$ (typically
    including the X-ray band). While the asymptotic flux (at $T
    \gg T_0$) above $\nu_c$ is unaffected by the density jump, the
    fluctuations in $\alpha$ are {\it comparable} to those below
    $\nu_c$.
\end{itemize}
An overdense clump on top of a uniform density background:
\begin{itemize}
\item Only a fairly large clump ($l_{\rm cl} \gtrsim R/\gamma$) with a
sufficiently large density contrast ($a \gtrsim 10-10^2$) produces a
significant fluctuation in the light curve.

\item The effect of a clump on the light curve is significantly larger
when it is located along the line of sight, than at an angle of $\sim
1/\gamma$ from the line of sight.

\item The signature of a clump is most apparent at $\nu < \nu_m <
\nu_c$.

\item Above $\nu_m$ a clump can actually cause a small dip in the light
curve, while below $\nu_m$ it causes a (larger) bump.

\end{itemize}
For a spherical density jump our conclusions are based on accurate
results, while in the case of a density clump we obtain only an
approximate upper limit for its effect on the light-curve.  Therefore,
our results for density clumps should be taken only as rough
guidelines. Our main results remain valid also when the observed
frequency is close to a spectral break frequency around the time of
the density jump.

Our conclusions are very different from those of previous works, which
predicted a significant optical rebrightening, and rather sharp
features in the afterglow light curve. The main cause for this
difference is our careful consideration of both the effect of the
reverse shock on the dynamics (which we find cannot be neglected even
when $a \sim 2$), and the arrival time of the photons to the
observer from different parts of the emitting regions. Both of these
effects tend to smoothen the light curve significantly.

Finally, we considered the implications of our results for the origin
of the fluctuations in the highly variable light curves of four GRBs
(3 long-soft GRBs and one short-hard GRB). We find that density
variations are unlikely to be the source of the fluctuations in any of
these bursts.

We thank Enrico Ramirez-Ruiz, Avishay Gal-Yam, Eran Ofek, Brad Cenko
and Pawan Kumar for useful comments. This research was supported by
a senior research fellowship from the Sherman Fairchild Foundation
(E. N.) and by US Department of Energy under contract number
DE-AC03-76SF00515 (J. G.).

\appendix

\section{A One Dimensional Special Relativistic Hydrodynamics and
  Radiation Code}\label{sec:code}

In order to obtain accurate light curves while relying on a minimal
number of approximations we use a one dimensional special relativistic
hydrodynamic code and combine it with an optically thin synchrotron
radiation module. This is the same code that was used before in
\cite{NP04}. We use a one dimensional hydrodynamic code that was
generously provided to us by Re'em Sari and Shiho Kobayashi. It is a
Lagrangian code based on a second-order Gudanov method with an exact
ultra-relativistic Riemann solver and it is described and used in
\cite{KPS99} and \cite{KS00}. On top of this code we have constructed
a module that calculates the resulting optically thin synchrotron
radiation. The code does not include the synchrotron self-absorption
or synchrotron self-Compton processes. The effect of the radiation on
the hydrodynamics is neglected. Below we describe the physics that is
included in the radiation module.

Having the full hydrodynamic evolution of the fluid (from the
hydrodynamic code) we first identify the time steps in which a given
fluid element is shocked by finding episodes of increase in its
entropy. The same fluid element can be shocked many times. Once a
fluid element is shocked all its electrons are assumed to be instantly
accelerated into a power-law energy distribution with an index $p$,
$dN/d\gamma_e \propto \gamma_e^{-p}$ for $\gamma_e > \gamma_{\rm
min}$. The energy in the electrons is taken as a constant fraction,
$\epsilon_e$, of the internal energy, and this condition determines
$\gamma_{\rm min}$ (it is assumed that $p > 2$). From this point on,
and until the same fluid element is shocked again, the electron energy
distribution decouples from the internal energy and evolves through
radiative cooling and adiabatic cooling or heating ($PdV$ work). The
magnetic energy in each fluid element is taken to be a constant
fraction, $\epsilon_B$, of the internal energy at all times.

One of the main difficulties in calculating the synchrotron radiation
at high frequencies is the short cooling time, which may be much
shorter than the hydrodynamic time steps. In order to overcome this
difficulty, we calculate the radiation during any hydrodynamic time
step analytically, in the following way. Immediately after a fluid
element crosses a shock, its initial electron energy distribution is
taken to be a power-law (with index $p$) between $\gamma_{\rm min}$
and $\gamma_{\rm max} = \infty$. The total emissivity of the fluid
element at a given frequency in its own rest frame, during a time
step, is obtained by integrating the spectral power of individual
electrons over the evolving electron distribution, where each electron
is tagged by the value of its initial Lorentz factor. The emission of
each electron is obtained by time integration over its instantaneous
emissivity\footnote{The synchrotron spectral power [erg/Hz/sec] of
each electron is approximated in the usual manner \citep{RL86}: $P_\nu
\propto \Theta(\nu_{\rm syn}-\nu)\,\nu^{1/3}$.}, which in turn depends
on the evolution of its Lorentz factor (and thus on its initial
Lorentz factor) during the time step. This evolution is calculated by
considering its radiative losses and its adiabatic cooling or
heating. In particular, we calculate the evolution of an electron with
initial Lorentz factor $\gamma_{\rm min}$ and of an electron with
initial Lorentz factor $\gamma_{\rm max}$. Their values at the end of
the time step are taken as the initial values for the next step in
which the initial distribution of electrons Lorentz factors is taken
again as a power-law between the new values of $\gamma_{\rm min}$ and
$\gamma_{\rm max}$. From the point where $\gamma_{\rm max}$ becomes
comparable to $\gamma_{\rm min}$ (within a factor of 2) the electron
energy distribution is taken as a delta function.

Since we use a one dimensional code in spherical coordinates, which
explicitly assumes spherical symmetry, each fluid element represents a
thin spherical shell. Once the rest frame spectral power of a fluid
element is calculated, we integrate over the contribution of this
shell to the observed flux at a given observer time and observer
frequency. This calculation takes into account the appropriate Lorentz
transformation of the radiation field and photon arrival time from
each point along the shell.

\end{document}